# Critical condition and transient evolution of methane detonation extinction by fine water droplet curtains

Jingtai Shi[a,b,c], Yong Xu[b], Wanxing Ren[a,c], Huangwei Zhang[b,*]

[a] *School of Safety Engineering, China University of Mining and Technology, Xuzhou, 221116, China*

[b] *Department of Mechanical Engineering, National University of Singapore, 9 Engineering Drive 1, Singapore 117576, Republic of Singapore*

[c] *Key Laboratory of Gas and Fire Control for Coal Mines, Xuzhou 221006, China*

## Abstract

Two-dimensional numerical simulations with Eulerian−Lagrangian method and detailed chemical mechanism are conducted to study the methane detonation propagation across a water curtain with finite thickness. The critical length of the water curtain with sprayed droplets is determined through parametric simulations with different water mass loadings and droplet sizes. The influence of water curtain length on the methane detonation is examined by the trajectories of peak pressure and time history of average heat release rate. The results indicate that the water curtain not only inhibit the incident detonation wave, but also prevent the detonation re-ignition after the incident wave is quenched. Moreover, unsteady response of gaseous methane detonation to water curtain are analyzed. The detonation re-initiation process behind the water curtain near the critical loading is also captured. In addition, mechanism of detonation inhibition with fine water droplets are discussed. It is found that energy and momentum exchanges start immediately when the detonation wave enters the water curtain area, but the mass transfer starts well behind the detonation wave due to the finitely long droplet heating duration. It is shown that the convective heat transfer by water droplets plays a significant role in quenching a detonation.

*Corresponding author. Tel.: +65 6516 2557; Fax: +65 6779 1459.
*E-mail address*: huangwei.zhang@nus.edu.sg.

## 1. Introduction

Methane explosions are common accidents in different applications, such as coal mines, metal mines, and natural gas pipelines. They may induce severe casualties and infrastructure damage. Effective suppression and mitigation technology are in high demand for methane explosion prevention and control [1]. Water spray mist is an ideal mitigant for gas explosion [2; 3]. It can absorb considerable heat from gas phase due to large heat capacity and latent heat of evaporation and release water vapor to dilute flammable gas [4; 5; 6].

There have been studies about methane combustion or explosion suppression with water sprays. Qin et al. [7] pointed out that explosion overpressure magnitude and rise rate, as well as blast wave propagation velocity, decrease with the concentration of ultra-fine water mist based on their experiments in a transparent rectangular cavity. Jing et al. [8] found that ultra-fine water mist above a critical concentration (160 g/m$^3$−800 g/m$^3$) can eventually lead to detonation extinction. Moreover, the droplet size is also shown to have a significant influence on gas explosion suppression. van Wingerden et al. [9] experimentally studied the influence of droplet size on pressure development during gas explosions in congested areas. They found that the effective explosion-mitigating water-spray systems are those generating very small droplets (less than 10 μm), because droplets of these sizes can fully vaporize in the reaction zone. The influences of initial droplet size and spray concentration on methane explosion inhibition were also investigated numerically by Song and Zhang [10]. Their results show that the explosion gas temperature can be reduced by 52.2%, when the initial spray concentration is 1.5 kg/m$^3$ and the initial droplet size is 150 μm. Thomas et al. [11] demonstrate that water sprays can significantly reduce the pressure and impulse levels generated during flame acceleration.

Since detonation is a confluence of chemical reaction zone and leading shock [12], studies are also available to understand how the water droplets affect the reaction and/or shock fronts in detonations. Liang et al. [13] found that addition of ultrasonically generated water mist significantly prolongs induced explosion time due to the combined effects of physical suppression and chemical inhibition. Jarsale et al. [14] found that the presence of water sprays considerably affects the cellular structure, based on their detonation experiments with ethylene-air mixtures. However, the ratio of the hydrodynamic thickness over the cell size is not sensitive to the water loading. Chauvin et al. [15] found the peculiar pressure evolution after the transmitted shock wave in two-phase mixture and they also measured the overpressures under different water spray conditions. Simulations were performed by Ananth et al. [16] to examine the effects of mono-dispersed fine water mist on a confined blast. It is found that the latent heat absorption is dominant for blast mitigation, followed by convective heat transfer and momentum transfer. Schwer et al. [17] modeled the interaction between the water mists and flow field generated by an unconfined explosion. They found that the water mists mitigate the shock-front pressure through energy and momentum extraction from drag and vaporization. In addition, they also observed that droplet size and mass loading play a secondary role to the total amount of water between the observer and the explosive blast. Recently, Watanabe et al. [18] observed that the dispersed water droplets significantly alter the hydrogen detonation flow field, and water droplet evaporation mainly occurs around 10 mm behind the leading shock. They also found that the cellular patterns of detonation in dilute water sprays are more regular than those of droplet-free detonations [19]. Xu et al. [20] predicted the critical curves of hydrogen detonation extinction with fine water sprays and it is found that the critical mass loading of water sprays increases with the droplet size. They also discussed the evolution of the chemical structure in the induction zone during a detonation extinction process.

However, studies on inhibition of methane detonations with water sprays are still very limited and hence the underlying controlling mechanism is not clear hitherto. In this study, two-dimensional simulations with Eulerian – Lagrangian approach are carried out to study methane detonation propagation across a water spray curtain with finite thickness. Detailed chemical mechanism is used for methane combustion, and the evaporating water droplets are tracked individually by solving their velocity, temperature, and size. The critical conditions for water mist to suppress methane detonation are explored, and the mechanism of methane detonation inhibition with water spray curtain is analyzed. The rest of the manuscript is structured as follows. The mathematical model is presented in Section 2, whilst the physical model is in Section 3. The simulation results and discussion will be given in Section 4, followed by the main conclusions summarized in Section 5.

## 2. Mathematical model

### 2.1 Gas phase

The governing equations of mass, momentum, energy, and species mass fraction are solved for compressible multi-component reacting flows. They respectively read

$$\frac{\partial \rho}{\partial t} + \nabla \cdot [\rho \mathbf{u}] = S_{mass}, \tag{1}$$

$$\frac{\partial (\rho \mathbf{u})}{\partial t} + \nabla \cdot [\mathbf{u}(\rho \mathbf{u})] + \nabla p + \nabla \cdot \mathbf{T} = \mathbf{S}_{mom}, \tag{2}$$

$$\frac{\partial (\rho \boldsymbol{E})}{\partial t} + \nabla \cdot [\mathbf{u}(\rho \boldsymbol{E} + \boldsymbol{p})] + \nabla \cdot [\mathbf{T} \cdot \mathbf{u}] + \nabla \cdot \mathbf{j} = \dot{\omega}_T + S_{energy}, \tag{3}$$

$$\frac{\partial (\rho Y_m)}{\partial t} + \nabla \cdot [\mathbf{u}(\rho Y_m)] + \nabla \cdot \mathbf{s_m} = \dot{\omega}_m + S_{species,m}, (m = 1, \ldots M - 1). \tag{4}$$

In above equations, $t$ is time and $\nabla \cdot (\cdot)$ is the divergence operator. $\rho$ is the gas density, $\mathbf{u}$ is the velocity vector, and $T$ is the gas temperature. $p$ is the pressure, updated from the equation of state, i.e., $p = \rho RT$. $R$ is the specific gas constant and is calculated from $R = R_u \sum_{m=1}^{M} Y_m W_m^{-1}$. $W_m$ is the molar weight of $m$-th species and $R_u = 8.314$ J/(mol·K) is the universal gas constant. In Eq. (4), $Y_m$ is the mass fraction of $m$-th species, and $M$ is the total species number. $\boldsymbol{E} \equiv e + |\mathbf{u}|^2/2$ is the total non-chemical energy, and $\boldsymbol{e}$ is the specific sensible internal energy.

The viscous stress tensor $\mathbf{T}$ in Eq. (2) modelled by $\mathbf{T} = -2\mu[\mathbf{D} - \mathrm{tr}(\mathbf{D})\mathbf{I}/3]$. Here $\mu$ is the dynamic viscosity and follows the Sutherland's law [21]. $\mathbf{D} \equiv [\nabla \mathbf{u} + (\nabla \mathbf{u})^T]/2$ is the deformation gradient tensor. In Eq. (4), $\mathbf{s_m} = -D_m \nabla(\rho Y_m)$ is the $m$-th species mass flux. The mass diffusivity $D_m$ can be calculated through $D_m = k/\rho C_p$ with unity Lewis number assumption. $C_p$ is the constant pressure heat capacity of the gas mixture. Moreover, $\dot{\omega}_m$ is the production or consumption rate of $m$-th species by all $N$ reactions. In Eq. (3), the combustion heat release $\dot{\omega}_T$ is estimated as $\dot{\omega}_T = -\sum_{m=1}^{M} \dot{\omega}_m \Delta h_{f,m}^o$, in which $\Delta h_{f,m}^o$ is the formation enthalpy of $m$-th species.

## 2.2 Liquid phase

The Lagrangian method is used to model the dispersed liquid phase with a large number of spherical droplets [22]. The interactions between the droplets are neglected because we only study the dilute water sprays (droplet volume fraction being less than 0.1% [28]). Therefore, the governing equations of mass, momentum, and energy for individual water droplets read

$$\frac{dm_d}{dt} = -\dot{m}_d, \tag{5}$$

$$\frac{d\mathbf{u}_d}{dt} = \frac{\mathbf{F}_d + \mathbf{F}_p}{m_d}, \tag{6}$$

$$c_{p,d}\frac{dT_d}{dt} = \frac{\dot{Q}_c + \dot{Q}_{lat}}{m_d}, \tag{7}$$

where $m_d = \pi\rho_d d_d^3/6$ is the mass of a single droplet, and $\rho_d$ and $d_d$ are the droplet material density and diameter, respectively. $\mathbf{u}_d$ is the droplet velocity vector, $\mathbf{F}_d$ and $\mathbf{F}_p$ are the drag and pressure gradient force exerted on the droplet, respectively. $c_{p,d}$ is the droplet heat capacity at constant pressure, and $T_d$ is the droplet temperature. In this work, both $\rho_d$ and $c_{p,d}$ are dependent on the droplet temperature $T_d$ [23].

The evaporation rate, $\dot{m}_d$, in Eq. (5) is modelled through

$$\dot{m}_d = k_c A_d W_d (c_s - c_g), \tag{8}$$

where $A_d$ is the surface area of a single droplet, $k_c$ is mass transfer coefficient, and $W_d$ is the molecular weight of the vapor. $c_S$ is the vapor mass concentration at the droplet surface, i.e.,

$$c_S = \frac{p_{sat}}{R_u T_f}, \tag{9}$$

where $p_{Sat}$ is the saturation pressure and is obtained under the assumption that the vapor pressure at the droplet surface is equal to that of the gas phase. The droplet surface temperature $T_f$ is calculated from $T_f = (T + 2T_d)/3$ [24]. In Eq. (8), the vapor concentration in the bulk gas, $c_g$, is obtained from

$$c_g = \frac{p x_i}{R_u T_f}, \tag{10}$$

where $x_i$ is the vapor mole fraction in the bulk gas.

The mass transfer coefficient, $k_c$, in Eq. (8) is calculated from the Sherwood number $Sh$ [25]

$$Sh = \frac{k_c d_d}{D_f} = 2.0 + 0.6 Re_d^{1/2} Sc^{1/3}, \tag{11}$$

where $D_f$ in Eq. (11) is the vapor mass diffusivity in the gas phase [26], and $Sc$ is the Schmidt number of gas phase. The droplet Reynolds number in Eq. (11), $Re_d$, is defined as

$$Re_d \equiv \frac{\rho d_d |\mathbf{u}_d - \mathbf{u}|}{\mu}. \tag{12}$$

In Eq. (6), the Stokes drag $\mathbf{F}_d$ is modelled as [27]

$$\mathbf{F}_d = \frac{18\mu}{\rho_d d_d^2} \frac{C_d Re_d}{24} m_d (\mathbf{u} - \mathbf{u}_d). \tag{13}$$

The drag coefficient in Eq. (13), $C_d$, is estimated as [27]

$$C_d = \begin{cases} 0.424, & if \ Re_d \geq 1000, \\ \frac{24}{Re_d}\left(1 + \frac{1}{6} Re_d^{2/3}\right), & if \ Re_d < 1000. \end{cases} \tag{14}$$

Since the ratio of gas density to the water droplet material density is well below one, the Basset force, history force and gravity force are not considered in Eq. (6) [28]. Besides, the pressure gradient force $\mathbf{F}_p$ in Eq. (6) is from

$$\mathbf{F}_p = -V_d \nabla p. \tag{15}$$

Here $V_d$ is the volume of single water droplet.

The convective heat transfer rate $\dot{Q}_c$ in Eq. (7) is calculated from

$$\dot{Q}_c = h_c A_d (T - T_d). \tag{16}$$

Here $h_c$ is the convective heat transfer coefficient, estimated following Ranz and Marshall [25]

$$Nu = h_c \frac{d_d}{k} = 2.0 + 0.6 Re_d^{1/2} Pr^{1/3}, \tag{17}$$

where $Nu$ and $Pr$ are the Nusselt and Prandtl numbers of gas phase, respectively. In addition, $\dot{Q}_{lat}$ in Eq. (7) is the heat transfer due to latent heat of droplet evaporation.

Two-way coupling between the gas and liquid phases is considered through Particle-source-in-cell approach [16]. The source terms, $S_{mass}$, $\mathbf{S}_{mom}$, $S_{energy}$ and $S_{species,m}$ in gas phase equations (1)−(4), are calculated based on the droplets in the CFD cells

$$S_{mass} = \frac{1}{V_c}\sum_{i=1}^{N_d} \dot{m}_{d,i}, \tag{18}$$

$$\mathbf{S}_{mom} = -\frac{1}{V_c}\sum_{i=1}^{N_d} \left(-\dot{m}_{d,i}\mathbf{u}_{d,i} + \mathbf{F}_{d,i}\right), \tag{19}$$

$$S_{energy} = -\frac{1}{V_c}\sum_{i=1}^{N_d} \left(-\dot{m}_{d,i}h_g + \dot{Q}_{c,i}\right), \tag{20}$$

$$S_{species,m} = \begin{cases} S_{mass} & for\ H_2O\ species \\ 0 & for\ other\ species. \end{cases} \tag{21}$$

Here $V_c$ is the CFD cell volume, $N_d$ is the droplet number in one cell and $h_g$ is the enthalpy of water vapor at droplet temperature. The droplet hydrodynamic force work is not included in Eq. (20), which is 2-3 orders of magnitude lower than the convective heat transfer and water vapor enthalpy in dilute spray detonations [20].

### 2.3 Computational method

The governing equations of gas and liquid phases are solved with *RYrhoCentralFoam* [29], which is developed based on the fully compressible non-reacting flow solver *rhoCentralFoam* in OpenFOAM 6.0 [30]. It has been extensively validated and verified for detonation problems in gaseous and gas−droplet two-phase flows and successfully applied for various detonation or supersonic combustion problems [31; 32; 33; 34; 35; 36].

Second-order backward method is employed for temporal discretization and the time step is about $1\times10^{-10}$ s. The MUSCL-type Riemann-solver-free scheme by Kurganov et al. [37] with van Leer limiter is used for convective flux calculations in the momentum equations. Total variation diminishing scheme is applied for the convection terms in energy and species equations. Also, second-order central differencing scheme is applied for the diffusion terms in Eqs. (2)−(4). A reduced chemical mechanism (DRM 19) developed by Kazakov and Frenklach [38] is used for methane combustion, which contains 21 species and 84 reactions. Its accuracy in modelling detonation initiation and propagation has been systematically studied by Wang et al. [39], including the ignition delay time over a range of operating conditions.

For the liquid phase, the water droplets are tracked based on their barycentric coordinates. The equations, i.e., Eqs. (5)−(7), are integrated by first-order implicit Euler method. Meanwhile, the gas properties at the droplet location (e.g., the gas velocity and temperature) are calculated based on linear weighted interpolation. The source terms, i.e., Eqs. (18)−(21), are calculated for the gas phase equations in a semi-implicit source approach. More detailed information about the numerical methods for gas and liquid phases can be found in Refs.[20; 29].

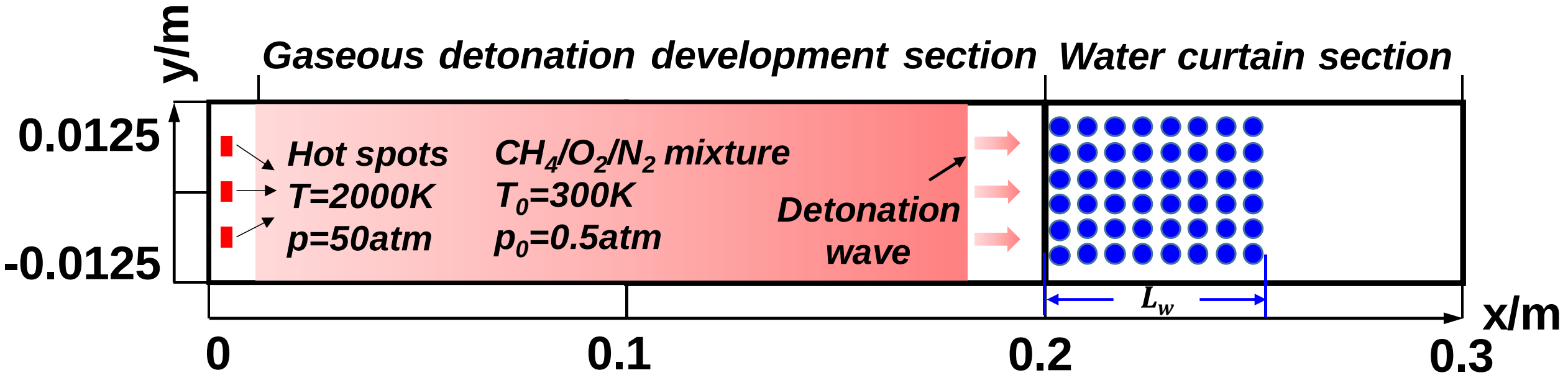

Figure 1 Schematic of the computational domain. Blue dots: water curtain with the length of $L_w$.

## 3. Physical model and numerical implementation

Two-dimensional (2D) methane detonation propagation across a water spray curtain with a finite thickness is studied in this study and the schematic of physical model and computational domain is shown in Fig. 1. The length (*x*-direction) and width (*y*-direction) are 0.3 m and 0.025 m, respectively. It includes gaseous detonation development section (0−0.2 m) and two-phase section (0.2−0.3 m), as marked in Fig. 1. The whole domain is initially filled with stoichiometric $CH_4/O_2/N_2$ (mole ratio of 1:2:1.88) mixture. The gas temperature and pressure are $T_0$ = 300 K and $P_0$ = 50 kPa, respectively. In the two-phase section, ultrafine water droplets are loaded to mimic the water curtain implementation for practical detonation explosion prevention, as demonstrated in Fig. 1. In this work, the water curtain always starts at $x$ = 0.2 and its streamwise length $L_w$ is varied to study how it affects the effectiveness for detonation inhibition. Note in passing that if $L_w = 0.1$, then the water droplets exist in the whole two-phase section. It should be highlighted that the water droplets are always dispersed along the entire width as shown in Fig. 1.

Cartesian cells are used to discretize the domain in Fig. 1 and the mesh cell size transitions from 50 μm ($x$ = 0−0.14 m) to 25 μm (0.14–0.3 m). To minimize the spatially variable resolution effects on detonation propagation, a refined area (0.14−0.2 m) with mesh size of 25 μm is included to connect the detonation development and two-phase sections. The resultant total cell numbers are 7,800,000. The Half-Reaction Length (HRL) estimated from the ZND (Zeldovich, von Neumann, and Döring) detonation structure of the stoichiometric $CH_4/O_2/N_2$ mixture is about 2,200 μm. Therefore, the resolution in the two-phase section is approximately 88 cells per HRL of C−J detonation. Considering the pronounced influence of the evaporating water droplets on the thermochemical composition in induction zone, actually over 88 cells can be expected within the HRL for spray detonations.

The detonation wave is initiated by three vertically placed hot spots (2,000 K and 50 atm) at the left end (see Fig. 1). For all the gas-liquid detonation modelling in this work, a consistent initial field with propagating detonation wave at about $x$ = 0.196 m (i.e., slightly before the two-phase section) is used. The upper and lower boundaries of the computational domain in Fig. 1 are periodic. For the left boundary ($x$ = 0), the *wave transmissive* condition in OpenFOAM is enforced for the pressure, whereas the zero-gradient condition for other quantities. Zero-gradient conditions are applied for the right boundary at $x$ = 0.3 m.

Monodispersed spherical droplets are considered in the water curtain. The initial water mass loading $z$ = 0.1 − 1.0 and droplet sizes of $d_d^0$ = 2.5 − 10 μm will be studied. Here the mass loading $z$ is calculated based on the water mass to the gas mass in the actual gas-droplet areas. In OpenFOAM [30], a small thickness (out-of-plane direction in Fig. 1) of the computational domain is needed for 2D simulations (but no numerical fluxes are calculated in this direction). This thickness is relevant when we estimate the domain volume for mass loading and two-phase coupling (i.e., Eqs. 18−20). Moreover, the initial temperature, material density and isobaric heat capacity of the water droplets are 300 K, 997

kg/m$^3$ and 4,187 J/kg·K, respectively. Besides, the water droplets are assumed to be initially stationary (i.e., $\mathbf{u}_d = 0$), which is reasonable due to typically small terminal velocities of fine droplets [15; 40].

To reduce the computational cost, virtual parcel concept is used, and one parcel represents many water droplets with identical properties (e.g., temperature, velocity and size) [28]. In this work, the initial number of water droplets within each parcel is assumed to be 10, i.e., $n_p^0$ = 10, following our previous studies [20]. When the water droplet breakup occurs subject to strong aerodynamic force from the high-Ma flows, this number density per parcel, $n_p$, increases considerably, but the total number of parcel (hence number of the solved Lagrangian equations, i.e., Eqs. 7−9) does not change throughout the simulation. In this study, droplet breakup process is modelled with the compressible version of the Pilch and Erdman model [41; 42], which accounts for a range of liquid aerodynamic breakup regimes depending on different Weber numbers.

## 4. Results and discussion

### 4.1 Critical conditions for detonation extinction

The critical length of the water curtain with sprayed droplets ($L_{wc}$) is determined through parametric 2D simulations with different water mass loadings and droplet sizes. The results are shown in Figs. 2 and 3, as a function of water mass loading $z$ and initial droplet size $d_d$, respectively. Here detonation extinction is identified from the gradually decaying Heat Release Rate (HRR) as shown later (e.g., Fig. 4) and numerical soot foils. The critical curve (i.e., the solid lines in Figs. 2 and 3) is determined based on the critical water curtain length demarcating the detonation extinction (open symbols in Fig. 2) and propagation (solid symbols) cases. In Figs. 2 and 3, methane detonations are always quenched by the water curtains above the critical water curtain length.

As demonstrated in Fig. 2, for a fixed droplet size, the critical water spray curtain length $L_{wc}$ decreases monotonically with water mass loading. For instance, when $d_d^0$ = 5 μm, $L_{wc}$ ≈ 0.023 m with $z$ = 0.1. Nonetheless, when $z$ is increased to 1.0, it is reduced to only around 0.003 m. This trend is observed for all the three droplet diameters in Fig. 2. Nonetheless, the dependence of $L_{wc}$ on water mass loading $z$ becomes weaker, when the mass loading is large. It is found that for $d_d^0$ = 2.5 and 5 µm, when the mass loading is higher than 0.8, the critical water curtain lengths respectively approach the limiting values of 0.0003 m and 0.002 m. This phenomenon may also exist for $d_d^0$ = 10 µm, although we do not simulate larger loadings (beyond 1.0) for it. This is a practically meaningful finding, because it implies that continuously increasing the water mass loading cannot ensure a smaller curtain length and meanwhile the effectiveness does not increase accordingly. Moreover, these limiting values of curtain lengths increases with the droplet sizes. This may be because they are associated with the characteristic timescales of the gas-liquid interactions, such as the thermal and momentum relaxation times [28].

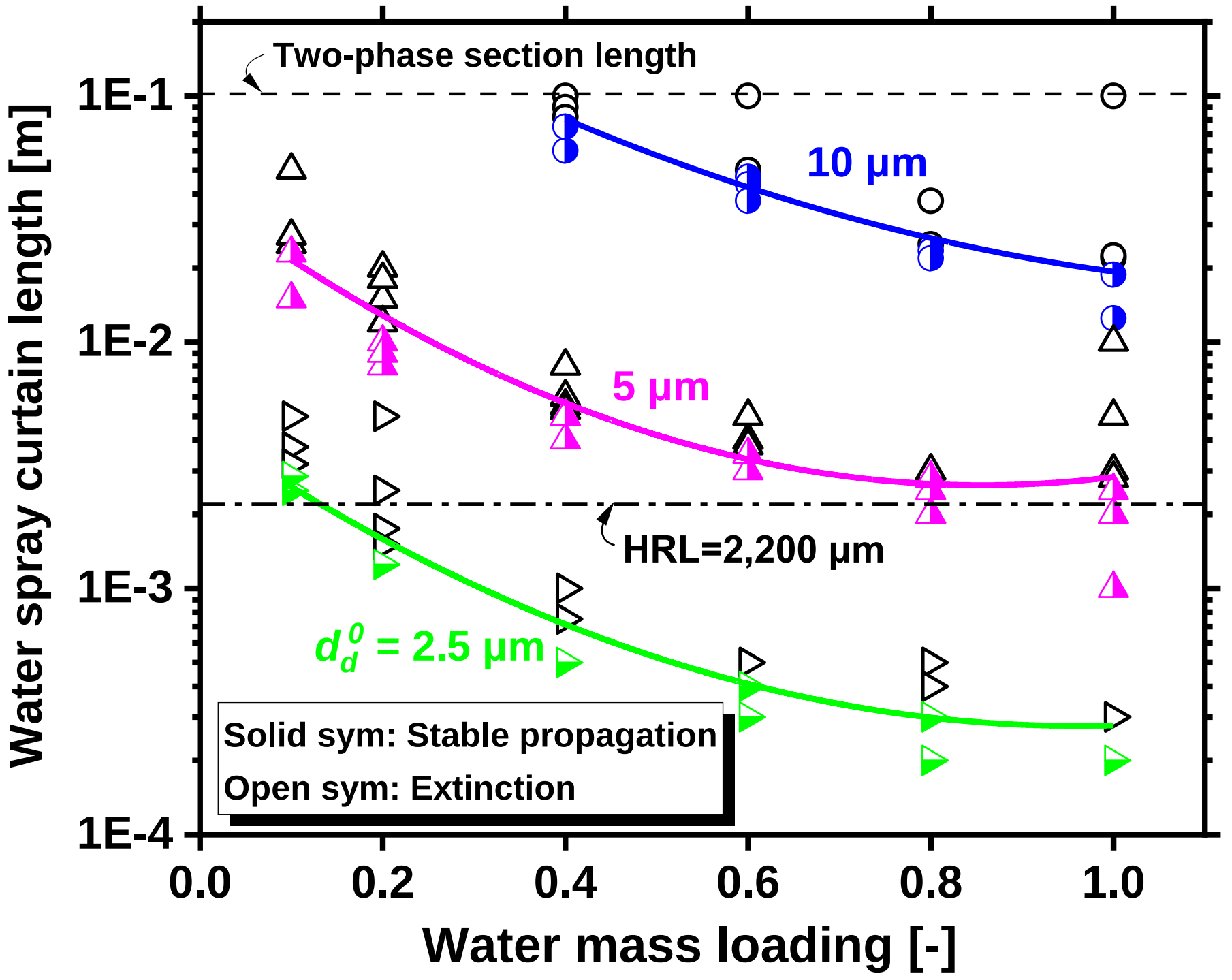

Figure 2 Critical length of water curtain as a function of water mass loading. Open symbol: detonation extinction; solid symbol: detonation propagation. Circles: 10 µm; upward triangles: 5 µm; rightward triangles: 2.5 µm.

In Fig. 3, for a fixed mass loading, the smaller the droplet size, the shorter the critical water curtain length. For instance, when the mass loading is 0.4, $L_{wc}$ is 0.075 m for the initial droplet size of 10 µm, whereas it is reduced to 0.00063 m for 2.5 µm. This is caused by the fast evaporation and heating rate of the smaller droplets and hence higher effectiveness in detonation inhibition. Besides, the specific area of smaller droplets also increases under the same mass loading conditions. Therefore, they can absorb more heat from the gas phase, more appreciably reduce reaction rate, and hence quench the detonation propagation. Also, due to the large amount of water vapor produced by droplets vaporization, the combustible gas is diluted, leading to earlier decoupling of detonation waves. For practical explosion hazard prevention measures, the results in Fig. 3 indicate that with a well sprayed water curtain can considerably reduce the water curtain length. This is of utmost importance because it can minimize the detonation-affected areas, thereby reducing the casualties and infrastructure damage.

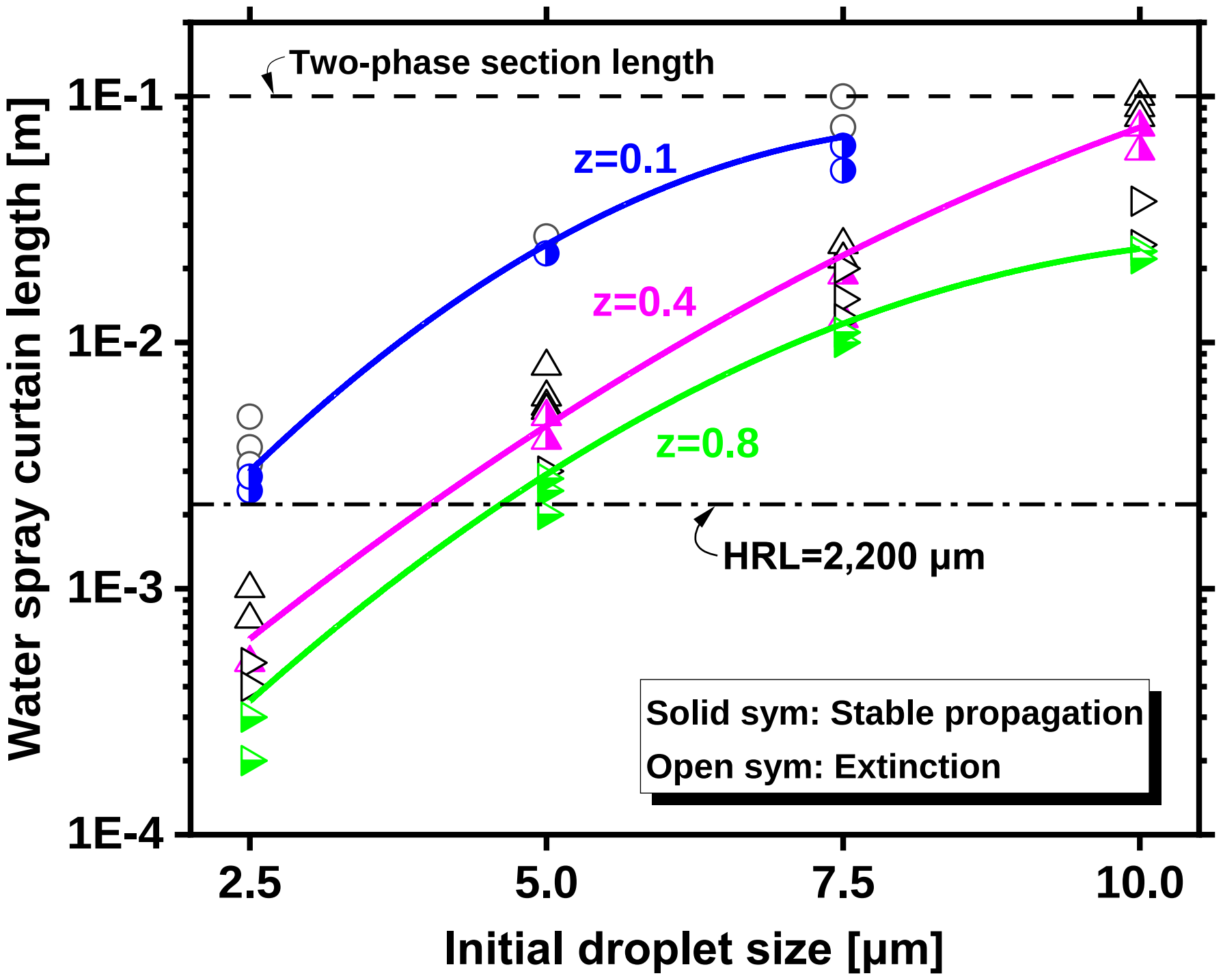

Figure 3 Critical length of water curtain as a function of initial droplet size. Open symbol: detonation extinction; solid symbol: detonation propagation. Circles: z = 0.1; upward triangles: 0.4; rightward triangles: 0.8.

### 4.2 Unsteady response of gaseous methane detonation to water curtain

To demonstrate how the gaseous methane detonations are influenced by the different water curtain lengths, the unsteady detonation evolutions crossing the water curtain will be discussed in this section. Four typical cases with different water curtain lengths are selected for detailed analysis here, i.e., $L_w$ = 0.0118, 0.0219, 0.0235 and 0.025 m. They have $d_d^0$ = 10 μm and $z$ = 0.8. The time evolutions of averaged HRRs are demonstrated in Fig. 4. Here the averaging is based on the domain of $x$ = 0.18−0.3 m. Each curve corresponds to complete propagation of detonation wave or shock wave from $x$ = 0.2 to 0.3 m, although the time durations in these cases are distinctive because of the different wave speeds. Note that the incident detonation wave arrives at the two-phase section at $t_{in}$ ≈ 3 μs ($t$ = 0 μs corresponds to the initial field for spray detonation simulation in which the detonation lies at $x$ = 0.196 m, as mentioned in Section 3). The result from $CH_4/O_2/N_2$ detonation without water curtain ($L_w$ = 0 m) is also added in Fig. 4.

When $L_w$ = 0.0118, the averaged HRR is generally steady but has some fluctuations. These fluctuations arise from the triple point collisions, which results in intermittently enhanced heat release. In this case, the evolution of HRR is relatively close to that of the water-free case with $L_w$ = 0 m. However, if $L_w$ is further increased, e.g., 0.0219 and 0.0235 m, the averaged HRR gradually decreases when it travels in the water curtain. Note that the time when the detonation wave leaves the curtain is marked as $t_{out}$ in Fig. 4. This signifies the gradually weakened detonative combustion in the gas phase by the water sprays. This tendency continues until 25 μs. After that, the averaged HRR suddenly increases and peaks around 35 μs, which is because detonation re-initiation occurs at the leeward side of the curtain (the unsteady process will be discussed at length later).

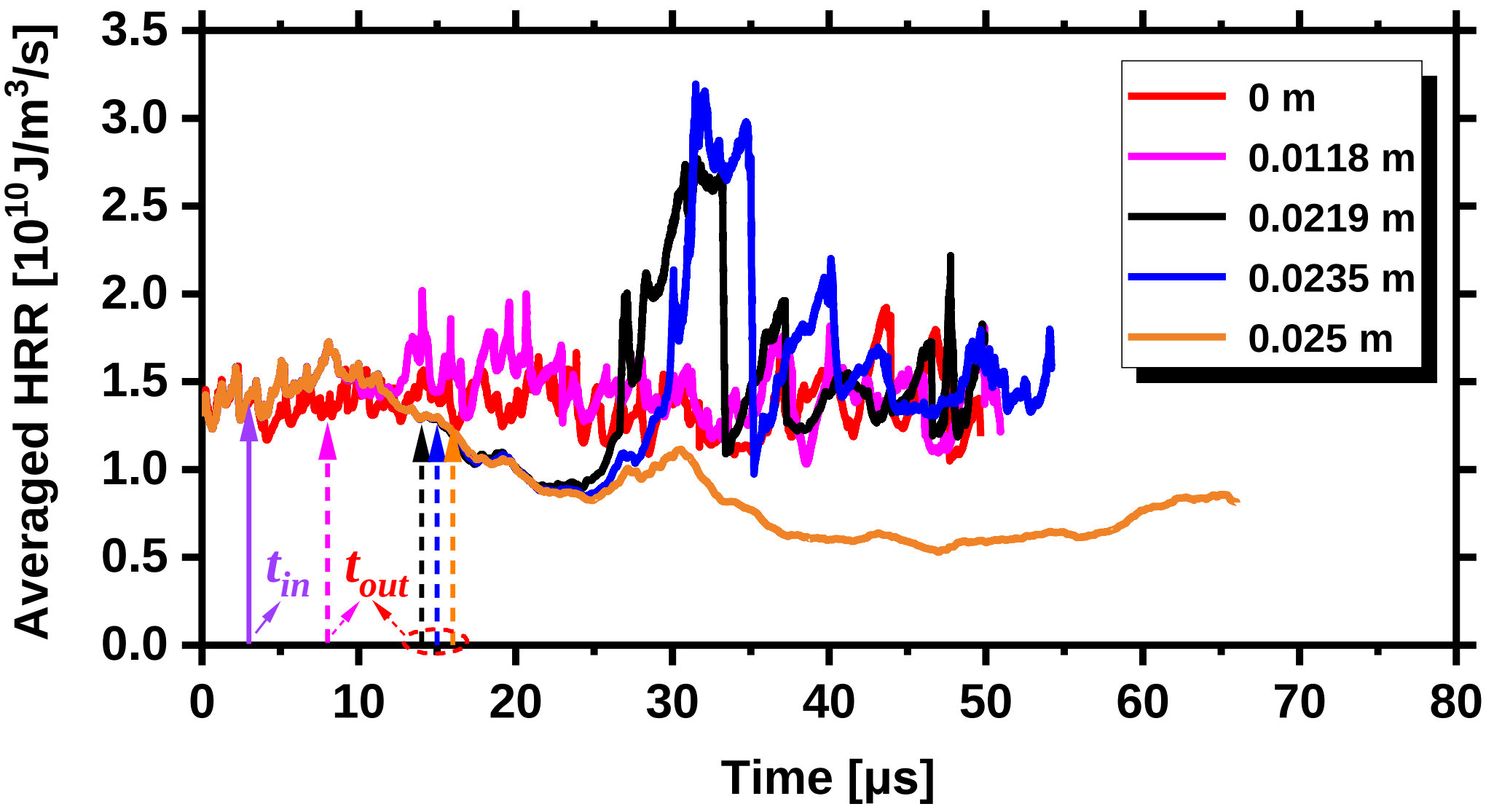

Figure 4 Time history of average heat release rate with different water curtain lengths ($L_w$). $d_d^0$ = 10 μm and z = 0.8. $t_{in}$ and $t_{out}$: the instant when the detonation enters and leaves the water curtain.

When the water curtain length is further increased to 0.025 m, as seen from Fig. 4, before 25 μs,

the HRR evolves similar to those with $L_w$ = 0.0219 and 0.0235 m. Nevertheless, no detonation re-initiation is observed in this case; instead, after 25 μs, the HRR gradually decreases, indicating the decoupling of the reaction front and leading shock front in methane detonation. It should be mentioned that the finite HRR stems from the residual reaction front running in the shocked combustible mixtures. How the travelling reaction front evolves in a shocked combustible gas is also a great concern for mitigation of potential secondary explosion in real accidents, since they may develop into a new detonation through deflagration-to-detonation transition when some induction factor (such as obstacle, turbulence and shock focusing [43; 44]) is present. Further studies on their longer evolutions after detonation extinction are merited as a future work.

Figure 5 further demonstrates the numerical soot foils of the cases in Fig. 4. They are recorded from the trajectory of maximum pressure location, normally from the triple points, when the detonation wave propagates. The initial locations of the water curtains are marked with red boxes. The result from water-free detonation is added in Fig. 5(a) for comparison. It can be observed in Fig. 5 that presence of water curtain considerably changes the cellular structures of stoichiometric $CH_4/O_2/N_2$ detonations. Specifically, when $L_w$ = 0.0118 m, the cell size within the water curtain is negligibly affected, through the comparisons between Figs. 5(a) and 5(b). However, beyond that, the cell size generally increases. For $L_w$ = 0.0219 m and 0.0235 m in Figs. 5(c) and 5(d), the detonation waves propagate a distance in the water curtain, and the triple points are considerably reduced at around $x$ = 0.022 m. Afterwards, the peak pressure trajectories quickly fade, signifying the decoupling of leading shock front from the followed chemical reactions in the course of gaseous detonation extinction. However, at around $x$ = 0.024 m, high-pressure spot arises in the middle of the domain width, resulting in X-shaped trajectories. This location is termed as Detonation Re-Initiation (DRI) point, as marked in Fig. 5. Downstream of the DRI, transverse detonation can be observed (see the dark trajectories extending from DRI), which burn the mixture in the lengthened induction zone due to the detonation failure. Clear cellular structures appear again, but the morphology changes considerably compared to those before the detonation interacts with the water curtains. One peculiar phenomenon is appearance of many fine cells inside the primary cells. This is also observed by Gamezo et al. [45] in their studies about marginal detonation near the propagation limits. They attribute it to the secondary pulsations generated by unstable overdriven parts of the leading shock front. These changes include larger cell sizes and secondary peak pressure trajectory. This re-initiation phenomenon is consistent with the sudden increased HRR in Fig. 4.

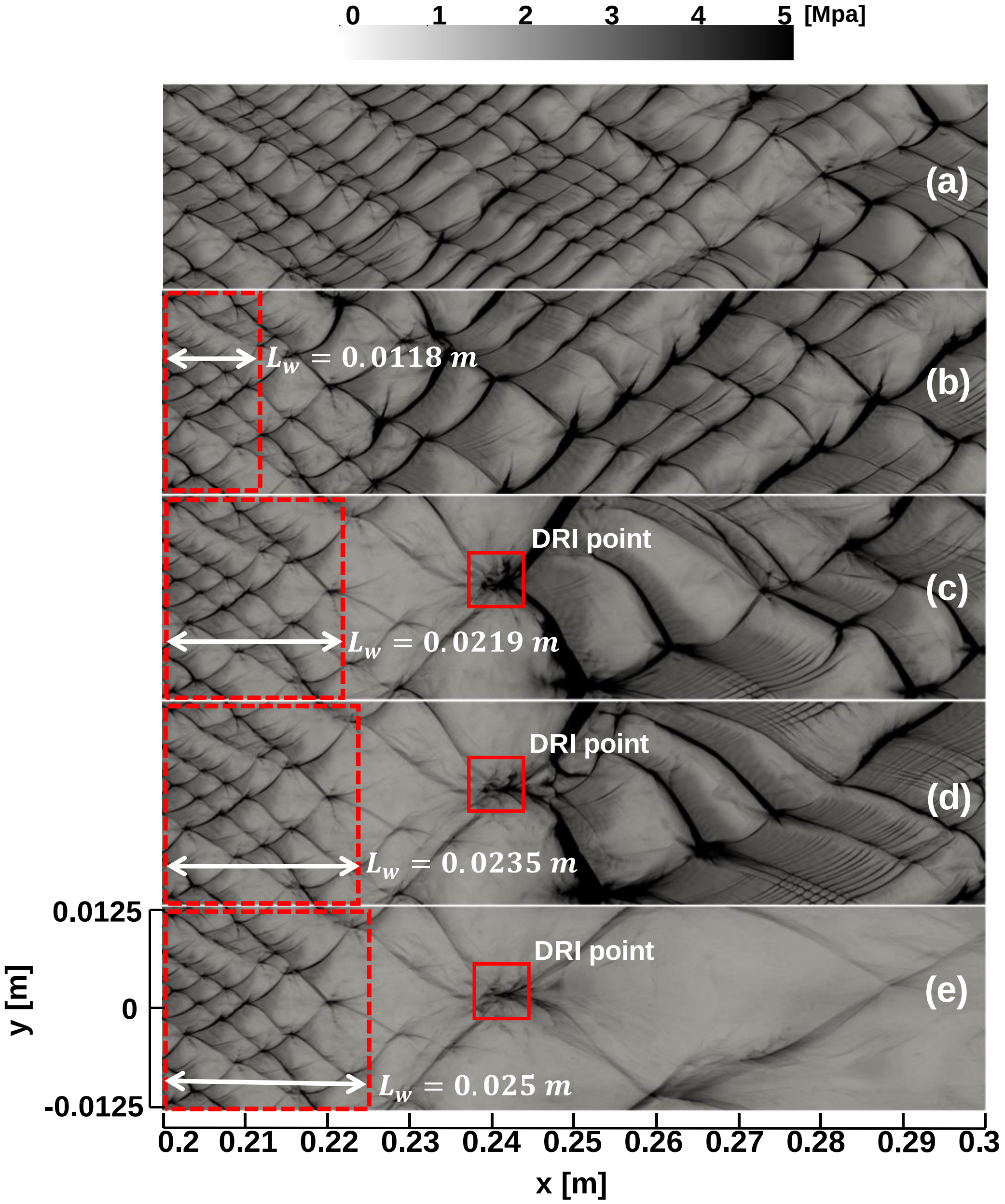

Figure 5 Peak pressure trajectory of detonation wave affected by water spray curtains with the length of: (a) $L_w$= 0 m, (b) 0.0118 m, (c) 0.0219 m, (d) 0.0235 m, and (e) 0.025 m. $d_d^0$ = 10 μm and z = 0.8. Dashed box: initial location of water curtain.

When the water curtain length is increased to 0.025 m, one can see from Fig. 5(e) that the peak pressure intensity gradually becomes weak after crossing the water curtain. The pressure slightly increases at the original DRI point but is generally much lower than the counterpart in Figs. 5(c) and 5(d). In this case, no detonation development is observed downstream. This difference results from the increased length of the upstream water curtain. Here one can find that only slight increments of $L_w$ leads to completely different outcomes of detonation evolutions. Therefore, caution needs to be exercised in practical water curtain implementations, because water curtain length close to the critical conditions is not recommended, although it may reduce the facility cost and complexity of the installation and arrangement. Based on Fig. 5, further increasing the water curtain length (> 0.025 m) would always quench the incident methane detonation.

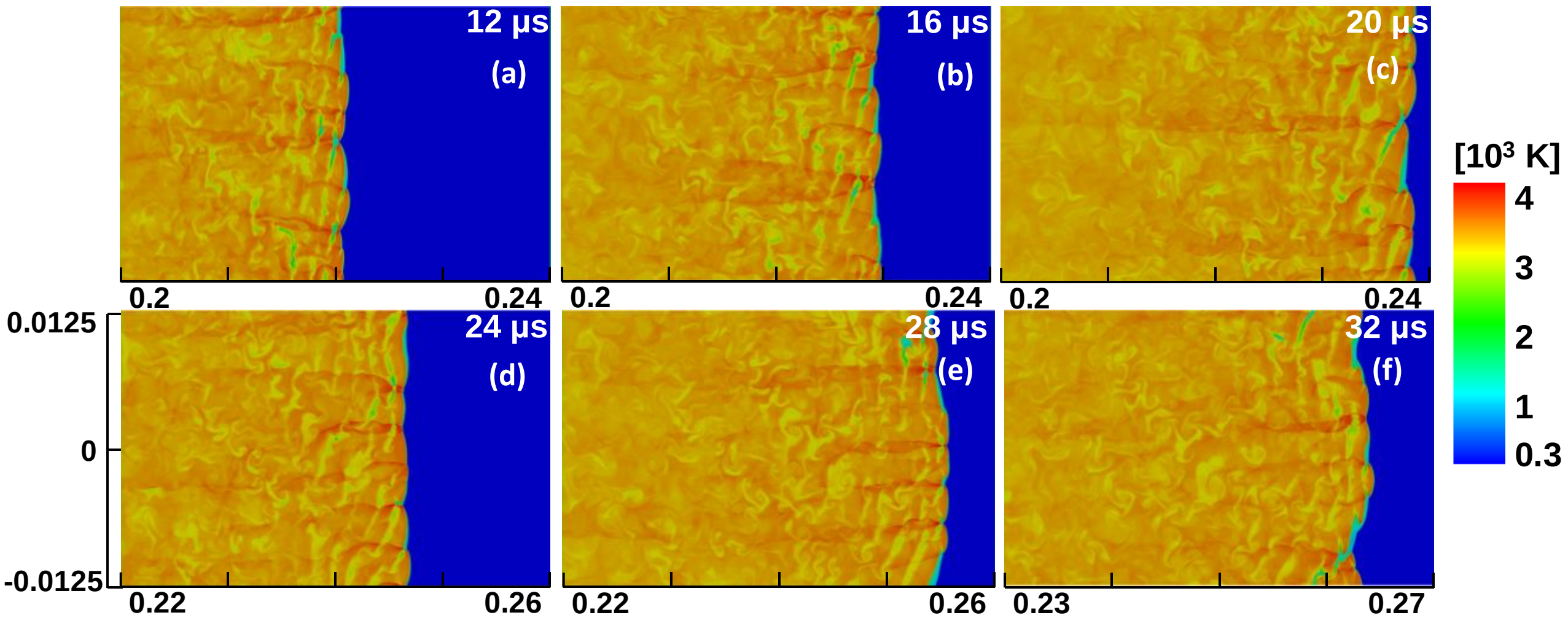

Figure 6 Time sequence of gas temperature distributions in a droplet-free methane detonation. Axis label in m.

Figure 6 shows the time evolutions of gas temperature in methane detonation without water curtain at six different instants, i.e., 12 – 32 μs. It is seen that the detonation can propagate stably, with multiple Mach stems, incident waves and transverse waves. Stripe structures of gas temperature are observed behind the detonation front, due to the compression by the propagating transverse shock waves. Figure 7 is the time sequence of gas temperature when the water curtain has a length of $L_w$ = 0.0118 m, corresponding to Fig. 5(b). Nine instants are visualized after the detonation wave crosses the water curtain (its boundary is approximately sketched with the left and right contact surface in Fig.

7). At 12 μs, the number of the Mach stem is appreciably reduced, leading to larger cell width, as seen from Fig. 5(b). The distance between the SF and RF is lengthened after the detonation wave crosses the water sprays. This can be clearly seen through comparing the results in Figs. 6(a) and 7(a). Continued propagation of the methane detonation wave is observed until the end of the domain at 44 μs, which is also manifested from the averaged HRR history in Fig. 4. Furthermore, the originally static water curtain (enclosed with the black lines) also moves along with the local high-speed flow field generated by the detonation wave, but the moving speed is obviously lower than the leading shock wave, due to the velocity relaxation time of the water droplets. Therefore, its direct inhibition influences on the travelling detonation wave quickly becomes weak, although it can still significantly lower the local gas temperature in the detonated areas.

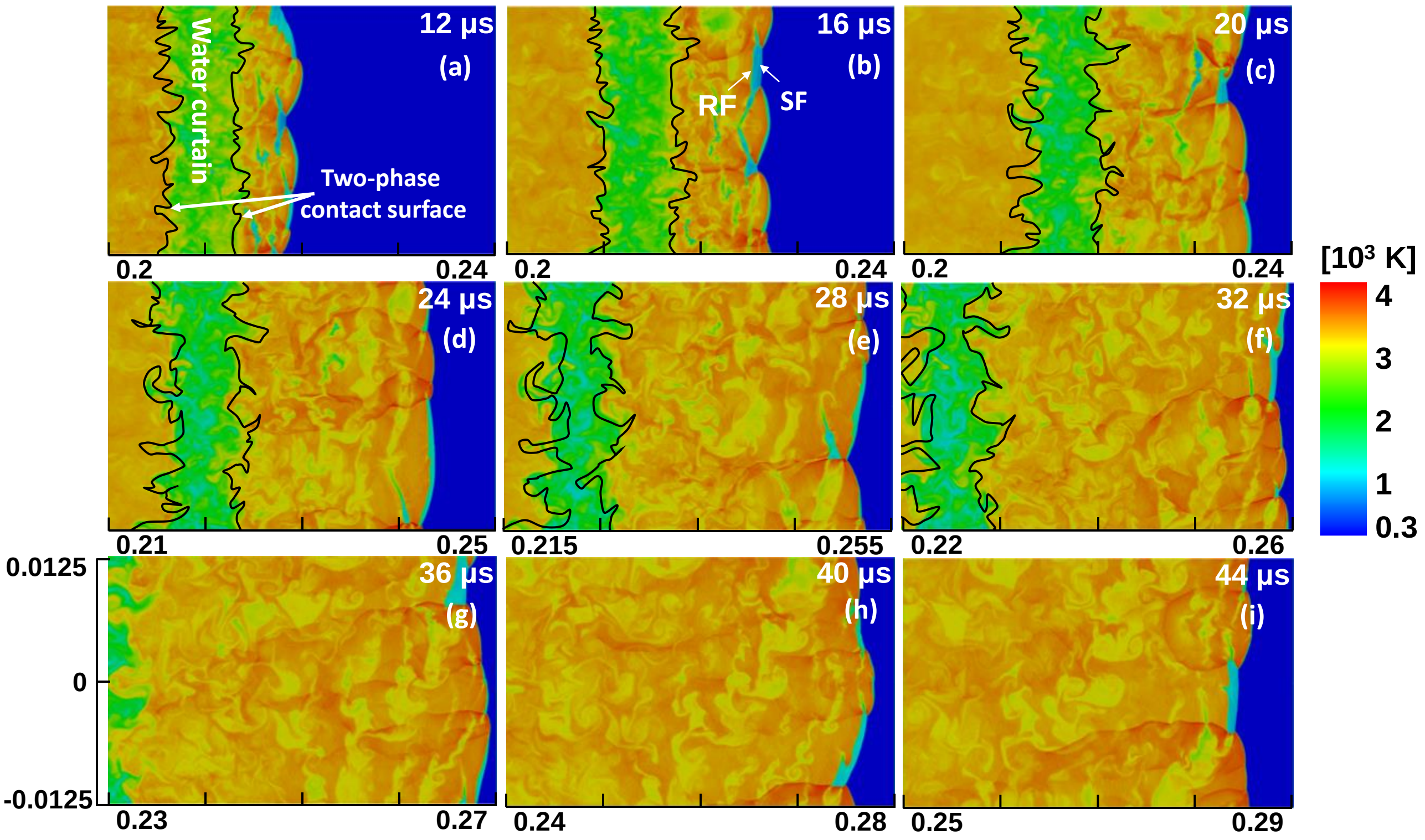

Figure 7 Time sequence of gas temperature distributions in a spray detonation with water curtain length of $L_w$ = 0.0118 m. $d_d^0$ = 10 μm and z = 0.8. Axis label in m. SF: shock front; RF: reaction front. Black lines: boundary of the water curtain.

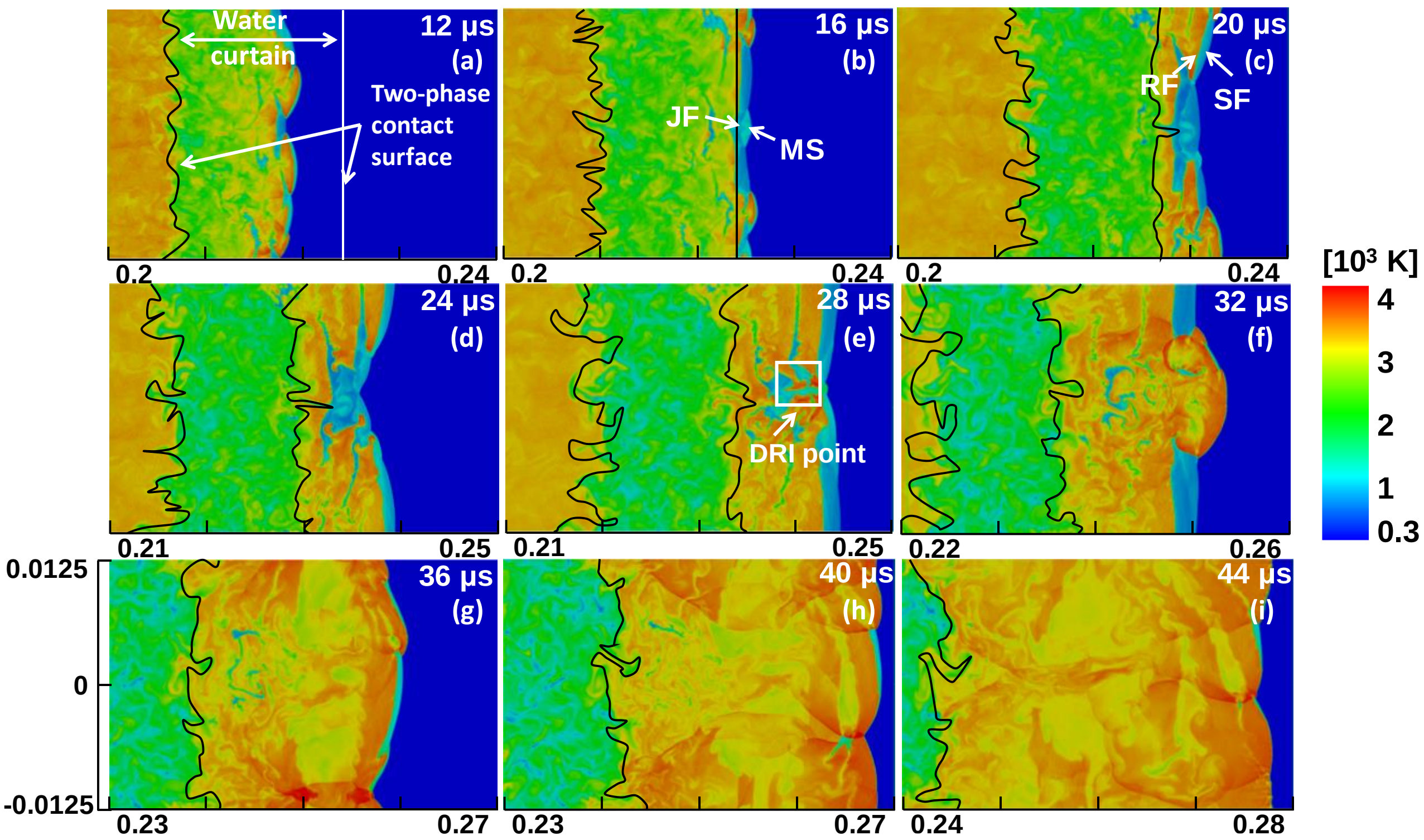

Figure 8 Time sequence of gas temperature distributions in a spray detonation with water curtain length of $L_w$ = 0.0235 m. $d_d^0$ = 10 μm and z = 0.8. Axis label in m. SF: shock front; RF: reaction front; MS: Mach stem; JF: jet flow. Black lines: boundary of the water curtain.

Figure 8 shows the counterpart results with a longer water curtain compared to that in Fig. 7, i.e., $L_w$ = 0.0235 m. Due to larger $L_w$, the residence time of the methane detonation wave inside the water curtain increases. For instance, at 12 μs, the detonation is still interacting with the dispersed fine water droplets. More pronounced effects on the detonation can be observed after the detonation crosses the curtain, such as at 16 and 20 μs. The induction zone between SF and RF is thickened, based on Figs. 6 and 7. Moreover, in the middle of the leading SF, no following reaction fronts are presented immediately (e.g., at 20 and 24 μs), indicating localized detonation extinctions. One can see from Figs. 8(e) to 8(f) that detonation re-initiation (see DRI point in Fig. 8e) occurs between 28 and 32 μs, leading to an overdriven Mach stem in Fig. 8(f). For better interpretation of the re-initiation process, the evolutions of gas temperature, pressure and HRR at 29, 30 and 31 μs are shown in Fig. 9. One can see that at 29 μs, the jet flows and triple points move towards each other. Their subsequent collision produces local high pressure and temperature. The gas in the induction area is ignited and two small forward hot jets are observed at 30 μs, and the coupling between them and leading SF results in local detonative combustion. This further evolves into a larger Mach stem, as seen from the results at 31 μs. Subsequently, from 36 μs to 44 μs, the detonation wave can continue propagating steadily, as found from Figs. 8(g)-8(i).

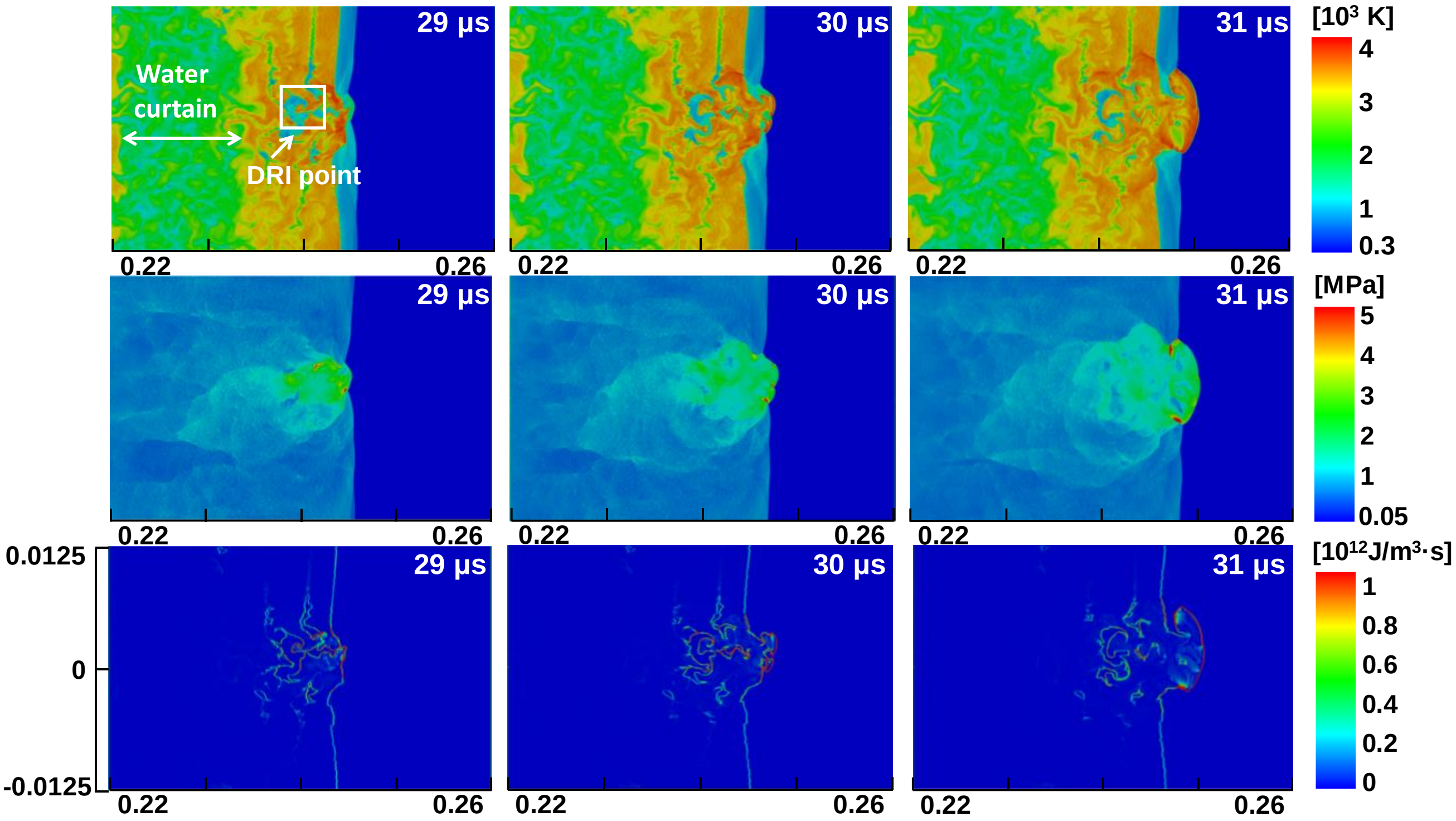

Figure 9 Distributions of (first row) gas temperature, (middle row) pressure, and (bottom row) heat release rate when the detonation is re-initiated after the water spray curtain. $d_d^0$ = 10 μm, z = 0.8, and $L_w$ = 0.0235 m. Axis label in m.

Plotted in Fig. 10 is the time sequence of the gas temperature at different time instants when the water curtain length is $L_w$ = 0.025 m. Before 28 μs, the evolutions of the detonation frontal structure are similar to the results in Fig. 8 with a shorter water curtain ($L_w$ = 0.0235 m). A forward reactive jet is seen at 32 μs in Fig. 10(f). Nonetheless, no mutual enhancement between the reactive jet and the leading curved SF is seen, and no detonative combustion develops. From 32 μs to 44 μs, although the reactive front grows considerably in the shocked $CH_4/O_2/N_2$ mixture, finite distance between the leading SF and RF always exists. Based on our numerical simulation results beyond 44 μs, no detonation re-initiation happens and therefore the detonation is deemed fully quenched after it passes the water curtain with $L_w$ = 0.025 m.

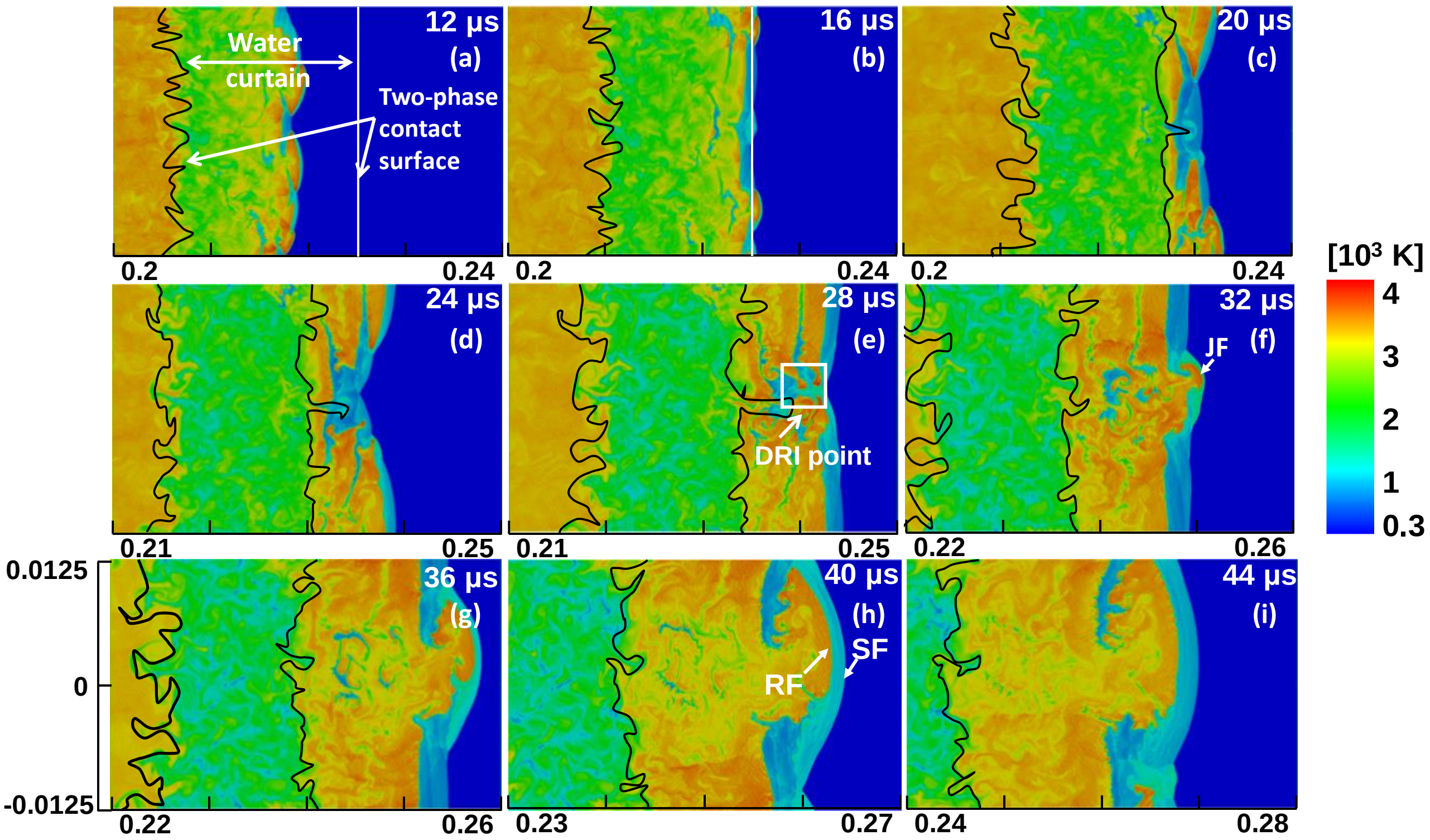

Figure 10 Time sequence of gas temperature when the detonation is quenched after the water spray curtain. $d_d^0$ = 10 μm, z = 0.8 and $L_w$ = 0.025 m. Axis label in m.

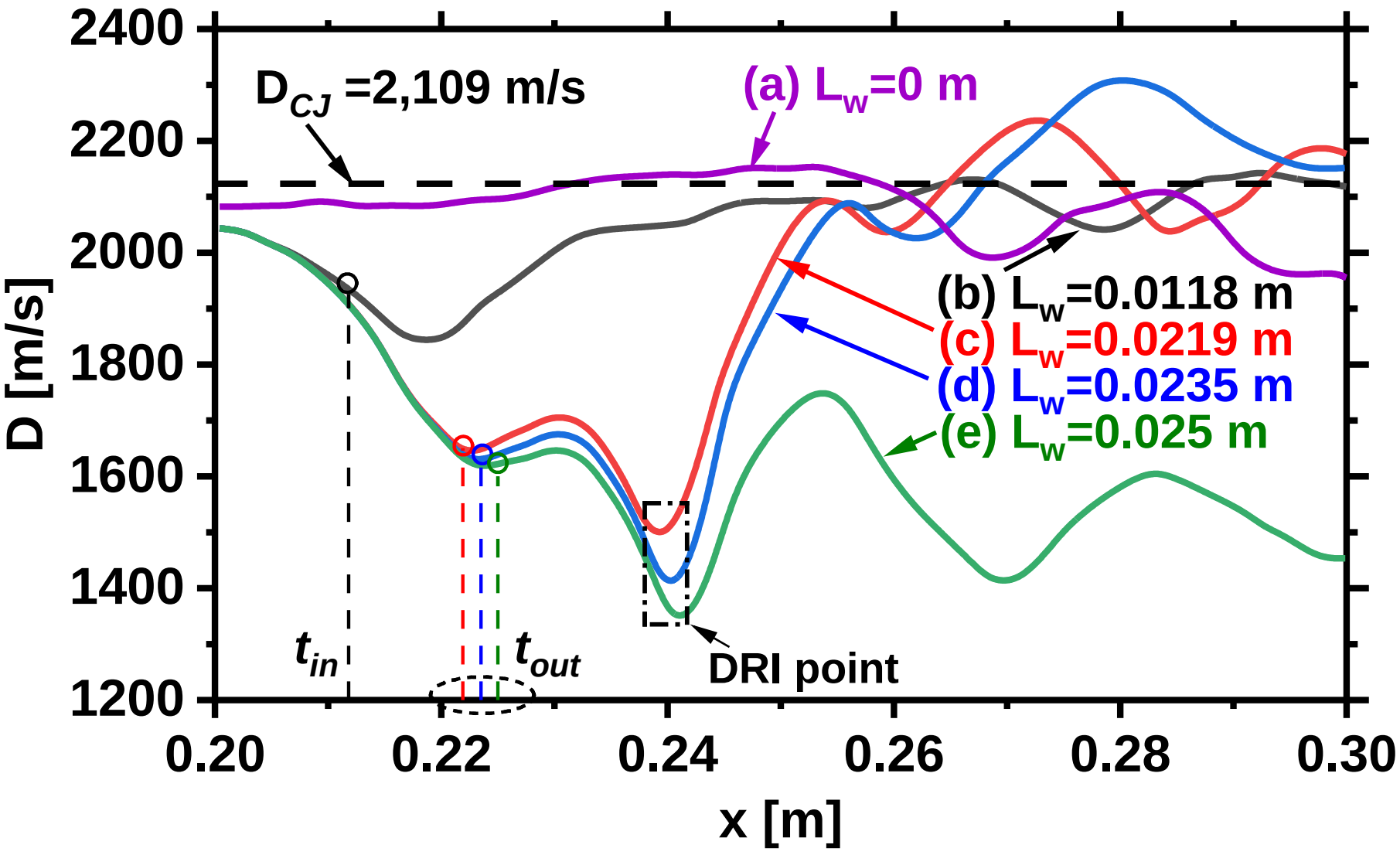

Figure 11 Spatial evolution of averaged leading shock propagation speed. $d_d^0$ = 10 μm and z = 0.8.

Figure 11 shows the spatial evolution of the averaged leading shock propagation speed $D$ in the foregoing two-phase methane detonations with different water curtain lengths. Note in passing that here the speeds are achieved with the segmented averaging method, to demonstrate the overall tendency of the detonation propagation behaviours, based on a timeseries (with a time interval of 1 μs) of leading shock position. For comparisons, we also add the C-J speeds $D_{CJ}$ of water-free $CH_4/O_2/N_2$ mixture for comparison, which is predicted by Shock and Detonation Toolbox [46] with the DRM 19 mechanism [38]. The detonation residence time in the curtain can be derived from $t_{in}$ and $t_{out}$ marked in Fig. 11.

As demonstrated in Fig. 11, the detonation propagation speeds of purely gaseous detonation are very close to the calculated C-J speed. However, the two-phase cases have speed deficits, relative to $D_{CJ}$. After the detonation wave passes the water curtain area, the speed of the detonation wave drops to the lowest value and then rises sharply near the DRI point. With increased $L_w$ from 0.0118 m to 0.025 m, the detonation waves are gradually reduced, which is particularly true after the detonation wave leaves the water curtain. Specifically, for $L_w$ = 0.0118 m, the speed fluctuates little near the C-J speed. For $L_w$ = 0.0219 m and 0.0235 m, the detonation wave re-ignites after the water curtain, and the speed is higher than the C-J speed at some locations, probably caused by the overdrive effects. When $L_w$ = 0.025, decoupling occurs after the detonation passes the water curtain. The wave speed decreases, well below the C-J value. Although the detonation wave is quenched, nonetheless, the blast wave degraded from the leading shock is still supersonic, with a speed of 1400-1600 m/s, which may still be disastrous for surrounding infrastructure and personnel. Therefore, how to quickly and effectively dampen the propagating blast wave necessitates further studies [17].

### 4.3 Mechanism of detonation inhibition with fine water droplets

The influences of water sprays on gaseous methane detonation are realized through mass, momentum, and energy exchanges between them. To reveal how these couplings play a role in quenching incident detonation, numerical experiments are performed. The base case is the extinction one with $L_w$ = 0.025 m (termed as case e hereafter), and their information is listed in Table 1. In case e1, the droplet evaporation model is switched off to rule out the mass transfer (water vapour addition) effects. In case e2, the droplet evaporation and heat transfer are not considered. Therefore, in this case, there is no mass and heat transfer, but only momentum transfer between the gas and droplet phases.

Table 1. Numerical experiments about detonation−droplet interactions

| case | Droplet evaporation | Convective heat transfer | Momentum transfer | Average shock speed [m/s] | |
|---|---|---|---|---|---|
| | | | | Water curtain | Whole domain (0.2−0.3m) |
| e | ✓ | ✓ | ✓ | 1,827 | 1,605 |
| e1 | × | ✓ | ✓ | 1,841 | 1,472 |
| e2 | × | × | ✓ | 1,960 | 2,027 |

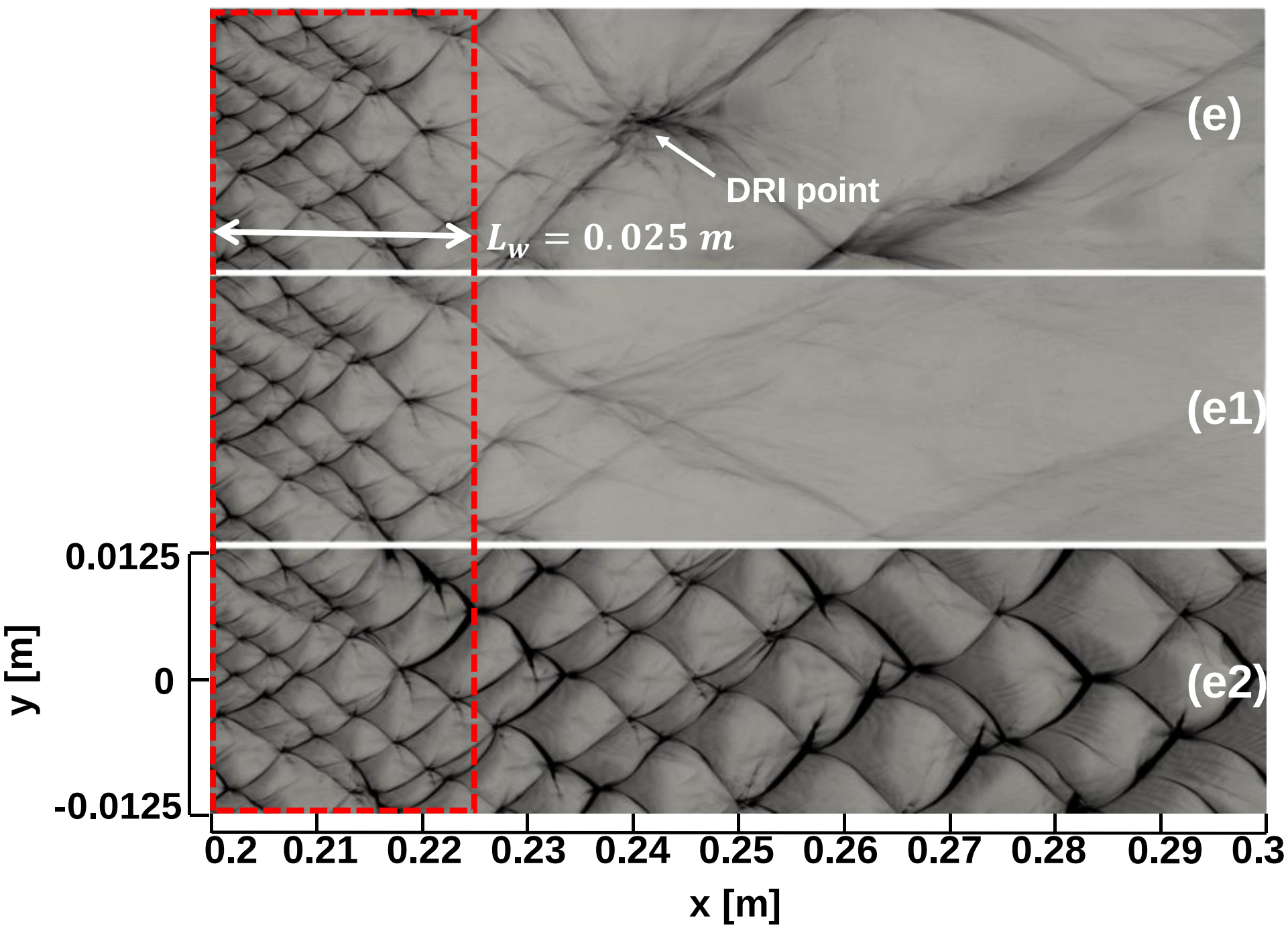

Figure 12 Peak pressure trajectory of propagating detonation wave in cases e (top), e1 (middle) and e2 (bottom). Dashed box: initial location of water curtain.

Figure 12 shows the trajectory of maximum pressures in cases e, e1 and e2. Note that the results for case e are identical to that in Fig. 5(e), where the detonation wave is decoupled after crossing the water curtain with a weak DRI point at around $x = 0.24$ m. For case e1, the detonation is also decoupled, but the re-initiation propensity is much lower, featured by weaker peak pressure values near the original DRI locus. One can see from case e2 that the detonation wave can propagate steadily after it passes the water curtain. The detonation cell size is increased to around 0.0125 m beyond the water curtain, indicating that the transverse waves are reduced and the detonation wave becomes more unstable. When the mass transfer and heat transfer models are de-activated in case e2, the propagation of detonation wave within the curtain is appreciably different from that in full case e. Moreover, the convective heat transfer has a more critical effect on detonation extinction, which can be found through the results in cases e1 and e2. From the three tests, we can have the following conjectures: (1) only momentum extraction by the gas phase (to accelerate the droplets) does not suffice to quench an incident detonation wave; (2) convective heat transfer is dominant in quenching a detonation wave; and (3) water vapour release from the droplets is shown to have limited influences on detonation wave dynamics. These will be further confirmed by the subsequent analysis in Figs. 15-18.

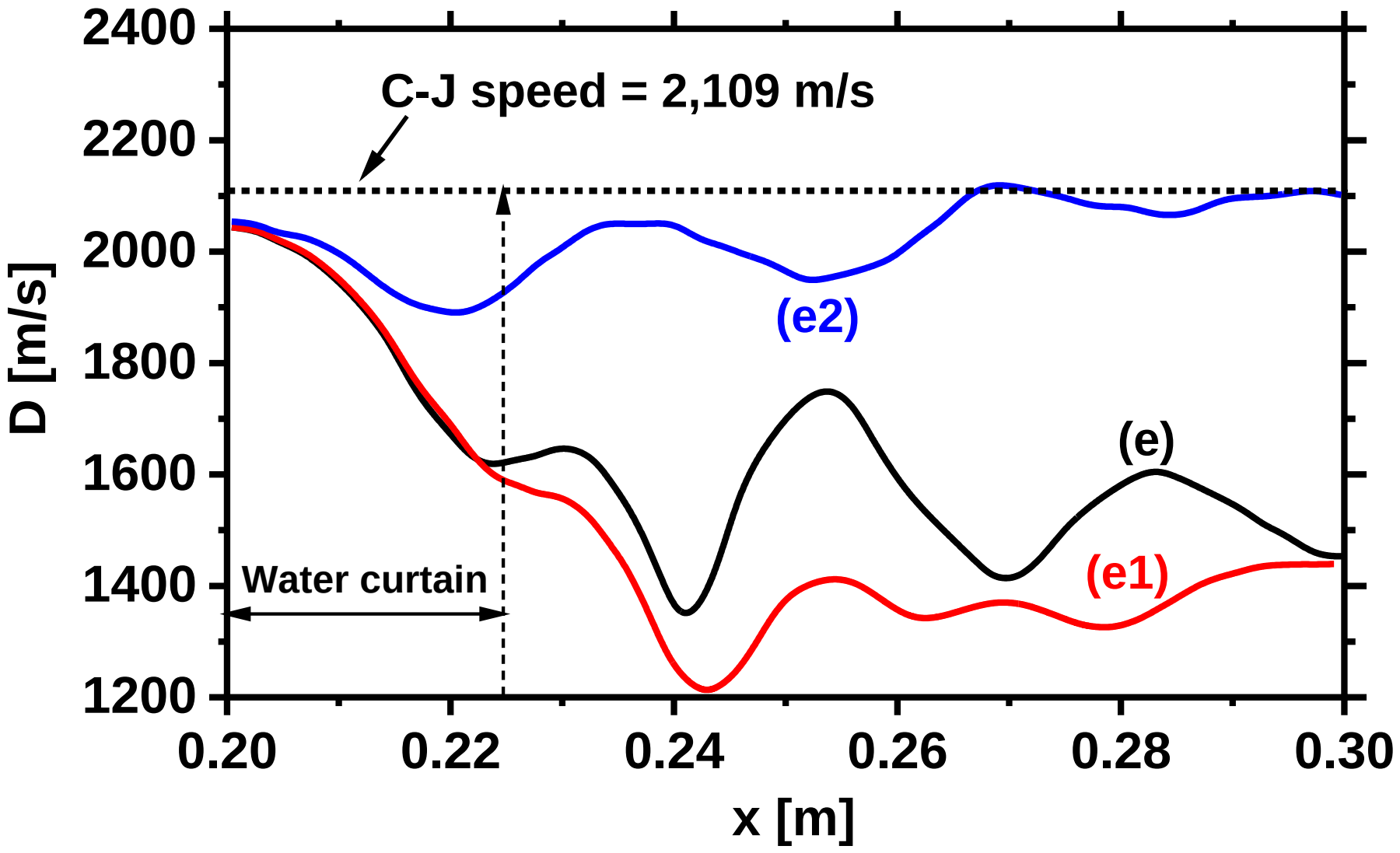

Figure 13 Spatial evolution of averaged leading shock propagation speed in cases e, e1 and e2.

Figure 13 further quantifies the evolutions of averaged shock propagation speed in cases e, e1 and e2. The calculation method is the same as for Fig. 11. In case e, the shock speed gradually decreases with some fluctuations when the detonation is decoupled after the water curtain. Compared with case e, the speed in case e1 is close to that of case e in the water curtain, and their average speed for crossing the detonation speeds are 1,827 and 1,841 m/s, respectively, as tabulated in Table 1. Beyond that, case e1 shows lower propagation speed compared to case e, corresponding to an average value (crossing the entire domain, see Table 1) of 1,472 m/s. This indicates weaker shock intensity due to absence of re-initiation when the droplet evaporation is not considered in the numerical experiments. For case e2, the shock is noticeably dampened in the water curtain, followed by a gradual increase to C-J speed after $x$ = 0.27 m. This is reasonable because the detonation wave is travelling in the gas-only mixture at these locations. Compared to the other two cases, the speed of case e2 is generally higher when only the momentum exchange is considered.

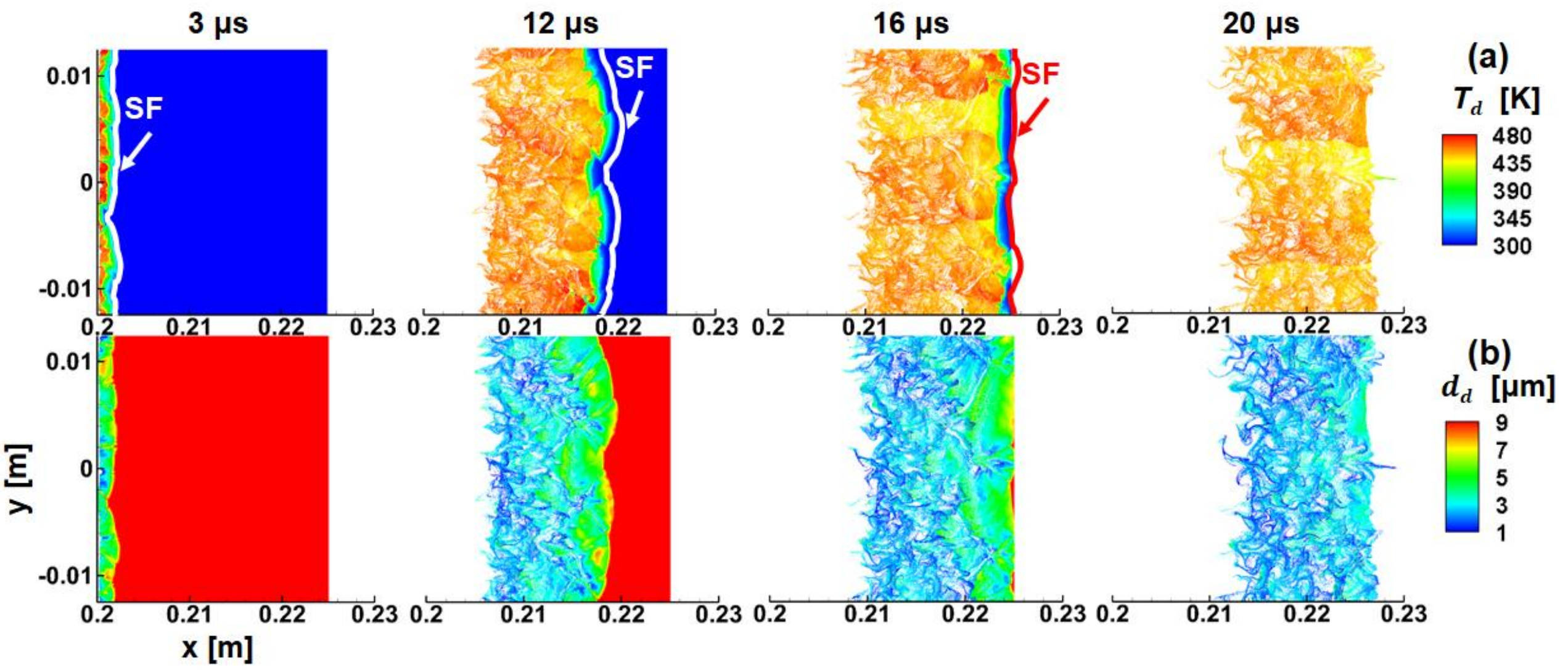

Figure 14 Time evolutions of water droplets colored by: (a) droplet temperature and (b) droplet diameter. Results from case e.

In the rest of this section, we will analyze the droplet and gas phase properties in case e to further analyze the mass, momentum, and energy exchanges. Figure 14 shows the time evolutions of water droplet temperature and diameter during the unsteady detonation extinction process. At 3 μs, the detonation wave arrives at the water curtain (the initial distribution is 0.2−0.225 m). In the shocked gas, the droplet temperature quickly rises, whilst the droplet size decays, due to aerodynamic fragmentation and evaporation. At 12 μs, it takes about 3 mm for the droplets to get heated towards its saturated temperature. The water curtain behind SF moves following the local detonated flows, and the evaporating droplets exist for about 12 mm behind the leading SF. The droplet diameter is reduced to less than 3 μm for most of the shocked water curtain area. The smaller droplet size would lead to smaller thermal and momentum relaxation timescale and hence is more conducive for two-phase interactions in terms of mass, momentum, and energy [11].

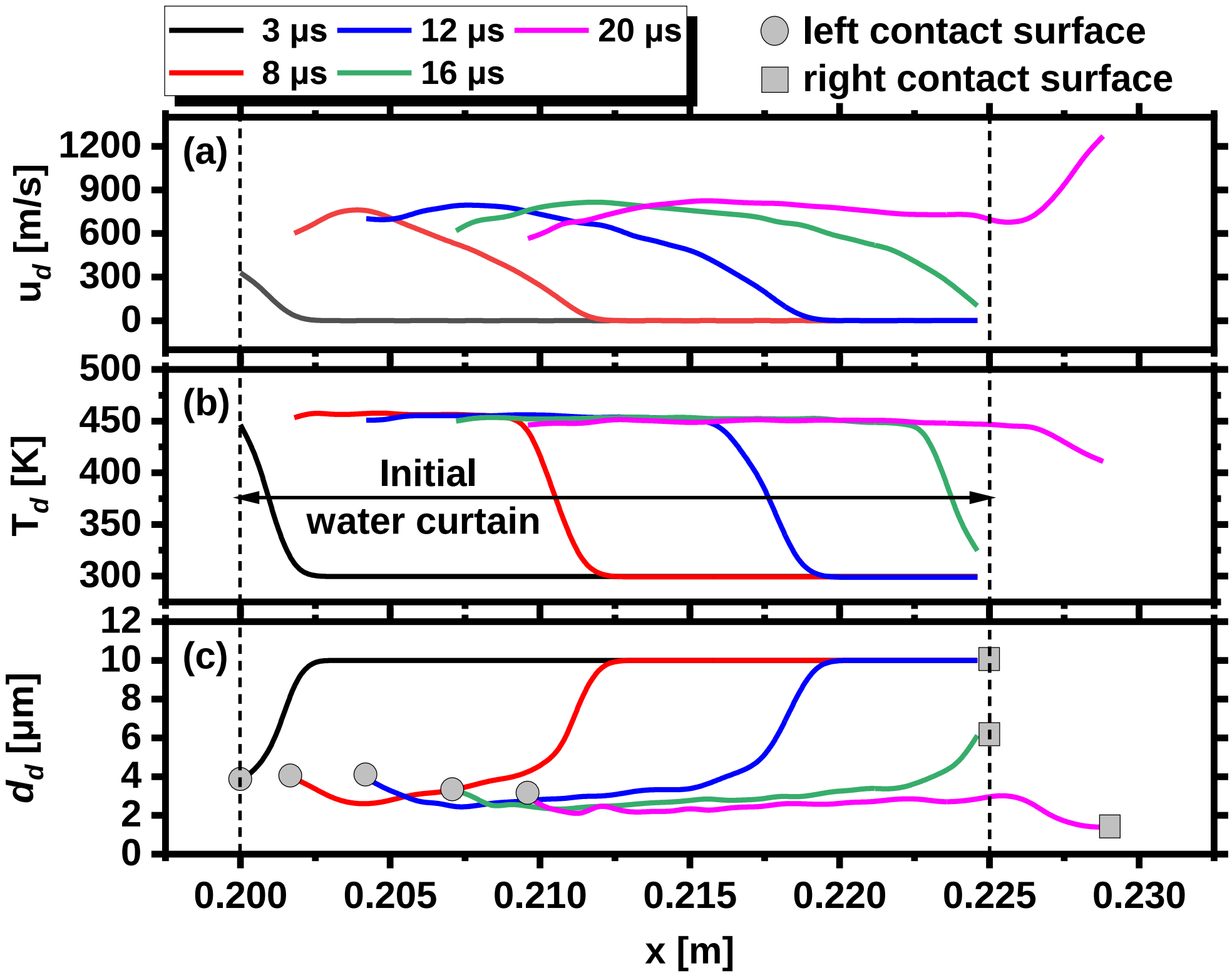

Figure 15 Spatial distributions of (a) velocity ($x$-component), (b) temperature, and (c) diameter of water droplets at five instants in case e.

Figure 15 shows the profiles of water droplet temperature, velocity, and diameter at five instants from case e. Note that these quantities are obtained through arithmetic averaging of the corresponding Lagrangian quantities (i.e., the results in Fig. 14) along the $y$-direction. The overall evolutions of these quantities are consistent with the results in Fig. 14. As observed from Fig. 15(a), the droplet velocity gradually increases due to acceleration by the detonation wave, and the peak values are around 800 m/s. These peak values are reached after finite distance after leading SF. It can be seen from Fig. 15(b) that the droplet temperature rises more quickly. It takes about 3 mm to 4 mm to reach a saturation temperature of approximately 450 K. Aerodynamic fragmentation leads to quickly reduced droplet diameter as shown in Fig. 15(c), from the initial value (10 μm) to around 3-4 μm. It can be seen from Fig. 15(c) that the droplets are gradually broken from 10 μm to about 3 μm.

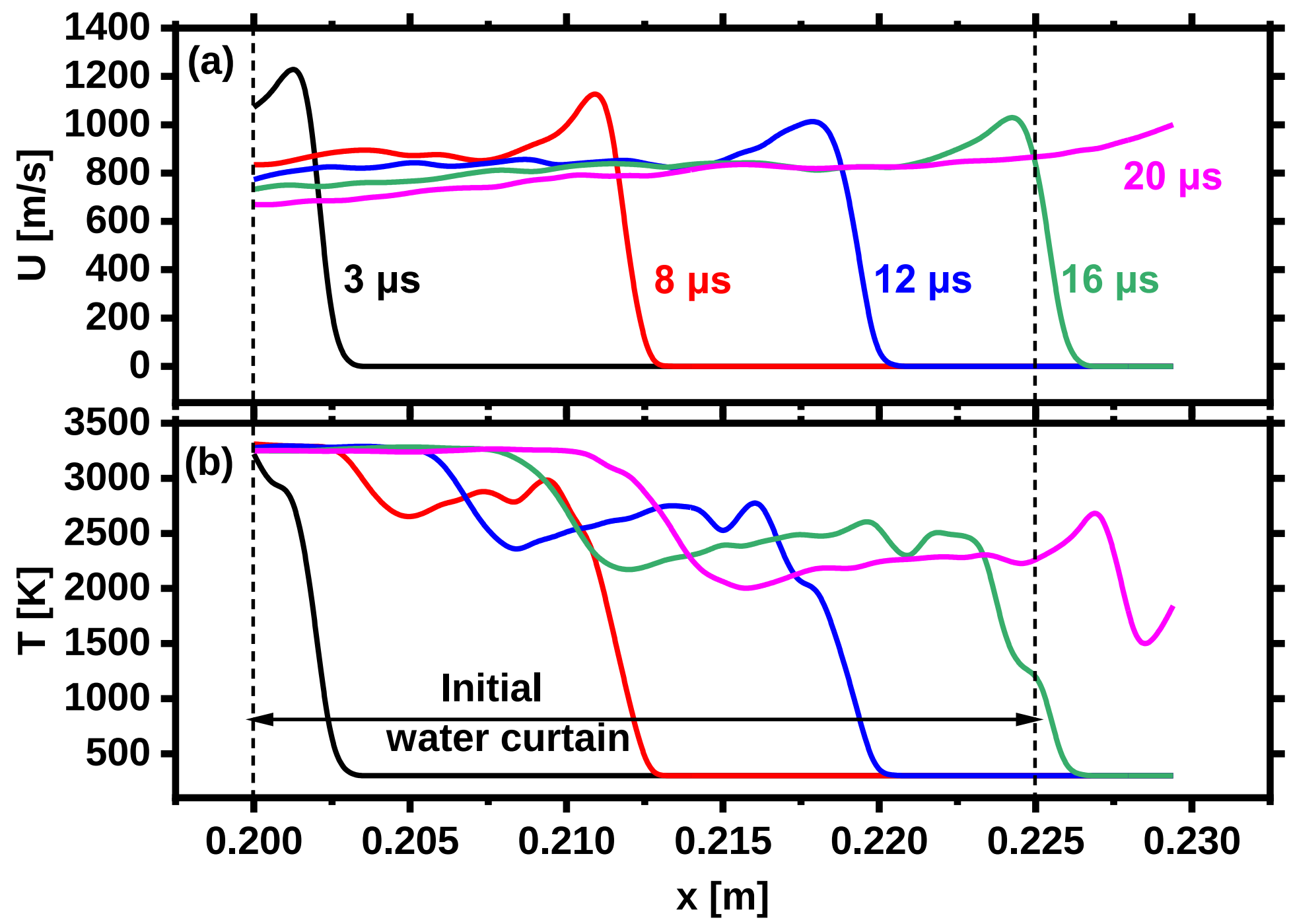

Figure 16 Spatial distributions of (a) gas velocity and (b) temperature at five instants in case e.

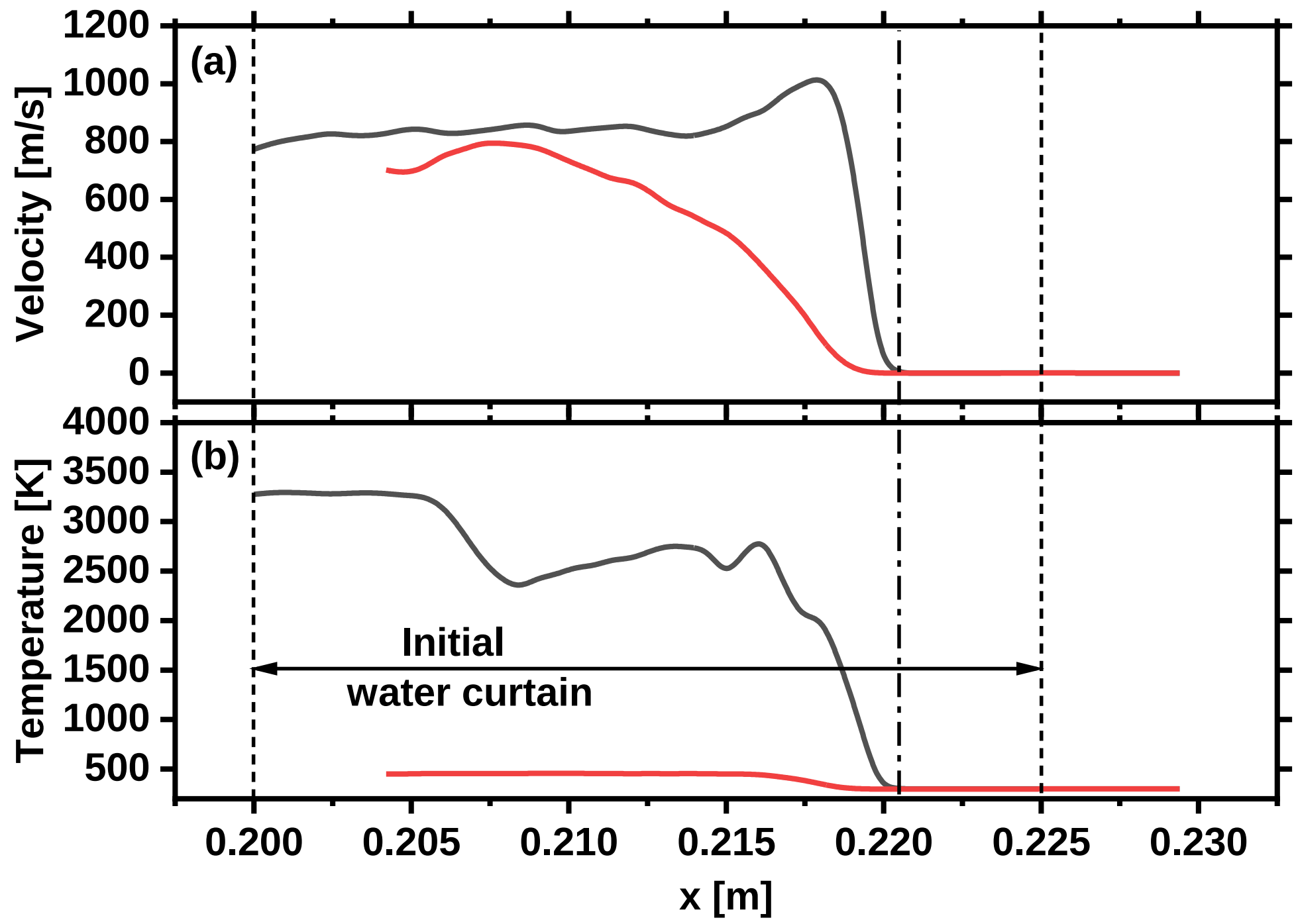

Figure 17 Spatial distributions of (a) velocity and (b) temperature at 12 μs in case e. Black lines: gas phase; red lines: liquid phase; dash-dotted lines: leading shock front.

Figure 16 shows spatial distributions of gas phase velocity and temperature at the same instants, which are obtained through density-weighted averaging along the *y*-direction. One can see from Fig. 16(a) that the averaged gas velocity grows quickly due to the arrival of the SF and decreases to about 800 m/s. As shown in Fig. 17(a), the gas and droplet velocities are close at around $x$ = 0.2075 m, where the droplet one peaks. This indicates a kinematic quasi-equilibrium location for gas and liquid phase. Nonetheless, further downstream, small velocity difference always exists. Meanwhile, it can be seen from Fig. 16(b) that the gas temperature also rises rapidly due to detonative combustion heat release, then have some finite variations between 2,000 K and 3,000 K, which spatially correspond to the droplet evaporation area. This can be clearly seen from Fig. 17(b). Considerable temperature difference exists in this area and therefore strong convective heat transfer for phase change would occur. Beyond the left contact surface, the gas temperature rises to a constant value of around 3300 K.

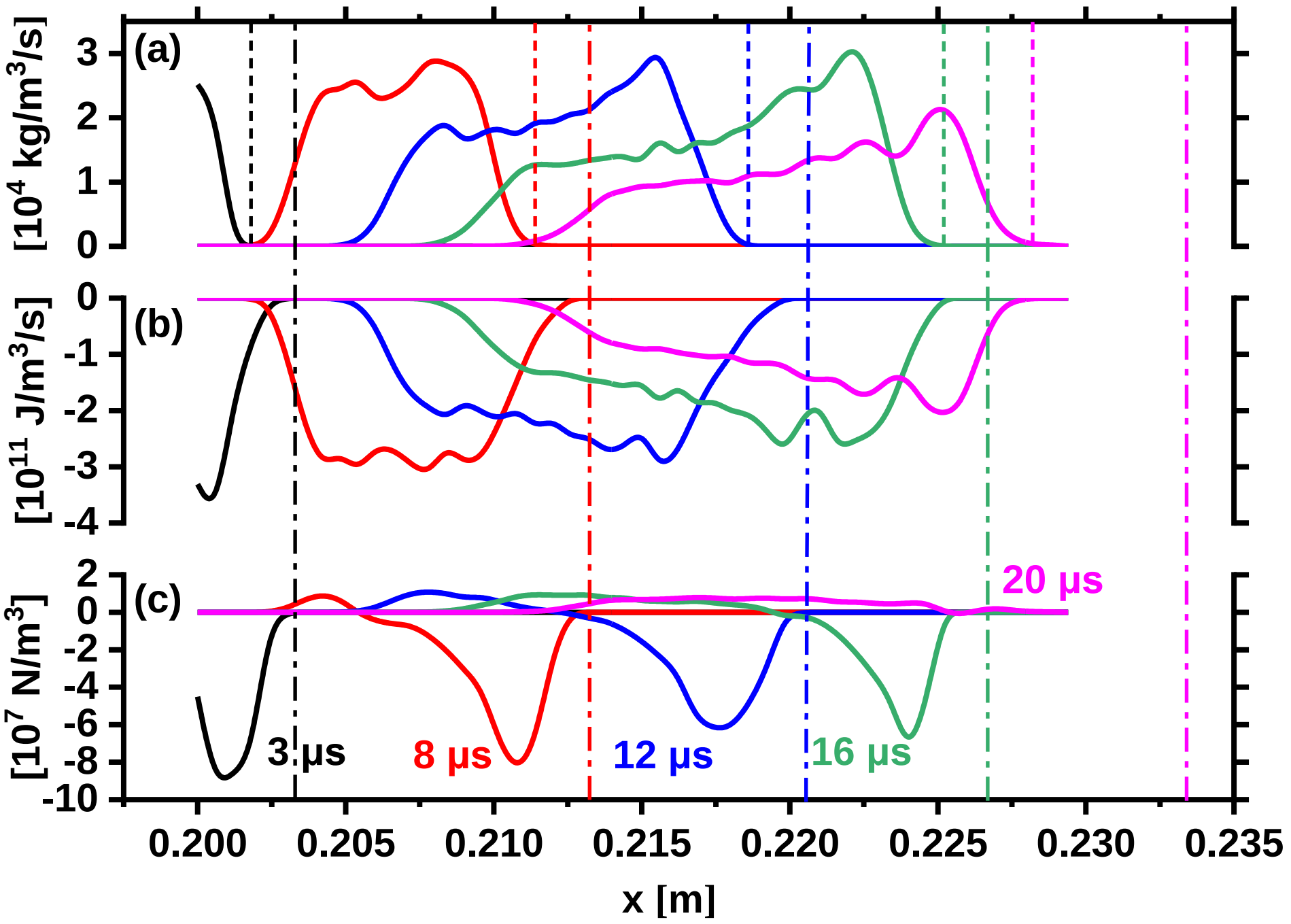

Figure 18 Profiles of averaged transfer rates of (a) mass, (b) energy and (c) momentum at five instants. Dash-dotted lines: leading shock fronts; dashed lines: reaction fronts.

The transfer rates of mass, momentum, and energy between the gas and liquid phases are presented in Fig. 18. The results are calculated from the density-weighted average interphase transfer rates ($S_{mass}$, $S_{mom}$ and $S_{energy}$ in Eqs. 18-20). A positive mass (energy and momentum) transfer rate indicates that the corresponding transfer is from liquid (gas) phase to gas (liquid) phase. The results in Fig. 18 correspond to the same instants in Figs. 15 and 16. It can be seen from Figs. 18(b) and 18(c) that the transfer of heat and momentum from gas to liquid phases proceed immediately behind the leading shock (dash-dotted line). Their magnitude gradually increases towards the downstream and reach the peaks around 2−5 mm after the shock. The heat transfer rate is relatively distributed due to the interphase temperature difference as shown in Fig. 17(b). Conversely, the momentum transfer rates in Fig. 18(c) are single-peaked. For instance, when $t$ = 12 μs, the maximum is observed at around 0.2175 m, which corresponds to the maximum gas phase and relatively low droplet velocity, as can found in Fig. 17(b). These tendencies about the energy and momentum exchanges can be found in all the shown instants.

However, the droplets start to vaporize around the RF due to increased temperature, and evaporation becomes significant well behind the detonation wave, i.e., the SF-RF complex (dashed lines in Fig. 18a). This is reasonable because of the finitely long droplet heating process, and implies that the water vapor from the water curtain may not have direct effects on the detonation structures, such as vapor dilution in the detonation induction zone. This also justifies the limited difference about detonation extinction between case e and case e1 in Fig. 12. Note that the droplet diameter is 10 μm. For other cases with finer diameters (5 and 2.5 μm), the same results are obtained.

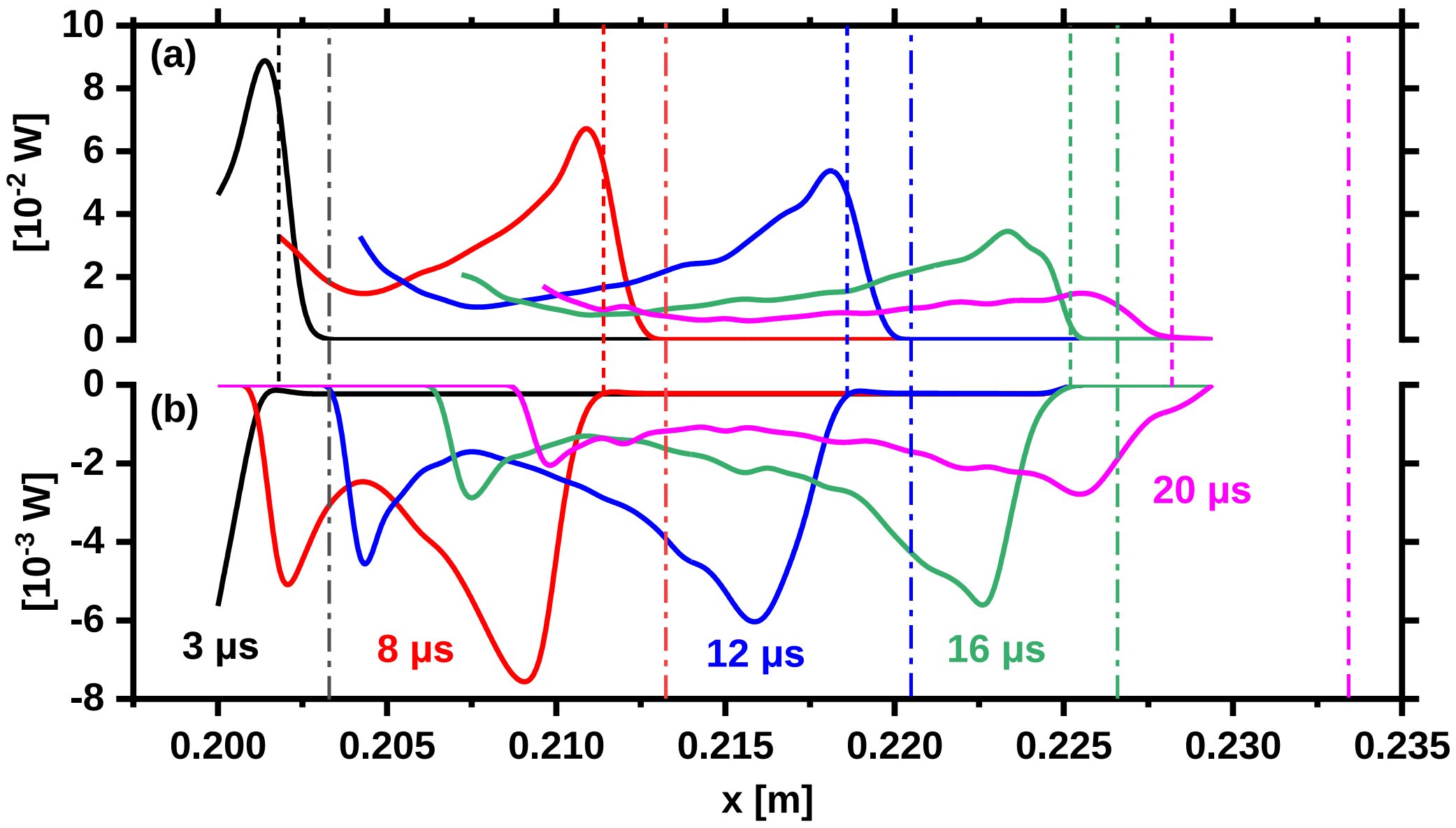

Figure 19 Profiles of averaged power from (a) convective heat transfer and (b) water vapor enthalpy from the Lagrangian water droplets. Dash-dotted lines: location of the shock fronts, dash lines: location of reaction fronts.

Based on the preceding analysis, it is known that heat transfer in quenching a detonation with fine water sprays plays an important role. Therefore, it would be helpful to compare the averaged power from the convective heat transfer and enthalpy of the added water vapor and the results are shown in Fig. 19. These two correspond to the mechanisms considered in the gas-liquid interaction, as shown in Eq. (20). The results are averaged from the Lagrangian droplet quantities along the *y*-direction. In general, for each instant, the power of convective heat transfer for the droplets is one order of magnitude higher than that for water vapor enthalpy. Since the droplet evaporation occurs well behind the RF and SF, the key mechanism for energy absorption near the detonation wave is convective heat transfer. This comparison further confirms conclusions from Figs. 12 and 13.

## 5. Conclusions

Extinction of methane detonation by fine water droplet curtains is studied with a hybrid Eulerian-Lagrangian method considering two-way gas−liquid coupling. Two-dimensional simulations with detailed chemistry for methane combustion are preformed. Different water mass loadings and diameters in the curtain are taken into consideration.

The critical length of the water spray curtain is determined through parametric simulations. The results show that the critical curtain length decreases monotonically with water mass loading. For a fixed mass loading, the smaller the droplet size, the shorter the critical water curtain length. When the water mass loading is beyond 0.8, the critical length approaches a constant value, and these constant values increase with droplet diameter.

The influence of water curtain length on methane detonation is examined by the trajectories of peak pressure and time history of averaged combustion heat release rate. The results indicate that the water curtain length has significant effects on unsteady detonation propagation behaviors. For a length smaller than the critical value, the detonation can cross the water curtain, but becomes more unstable. When the length is close to the critical value, decoupling of SF and RF occurs and re-initiation occurs behind the water curtain. When the length is above the critical value, the incident detonation wave can be quenched without re-initiation.

Moreover, unsteady response of methane detonation to the water curtain is studied. General features of gas phase and liquid droplets and detailed detonation frontal structures are well captured. Incident detonations in different water curtain lengths are discussed, about the evolutions of frontal structure and detonation propagation speed. Compared to the gaseous cases, the two-phase cases have pronounced speed deficit, relative to the dry mixture C-J speed. It is seen that with increased curtain length from 0.0118 m to 0.025 m, the detonation wave speed generally decreases. When the water curtain is 0.025 m, although the detonation extinction occurs, nonetheless, the blast wave degraded from the leading shock is still supersonic.

In addition, mechanisms of detonation inhibition with fine water droplets are discussed. It is found that energy and momentum exchanges start immediately when the detonation wave enters the water curtain area, but the mass transfer starts well behind the detonation wave due to the finitely long droplet heating duration. It is shown that the convective heat transfer by droplet heating plays a significant role in quenching a detonation.

## Declaration of Competing Interest

The authors have no conflict of interest.

## Acknowledgements

This work used the computational resources of ASPIRE 1 Cluster in The National Supercomputing Centre, Singapore (https://www.nscc.sg), and Fugaku Cluster provided by RIKEN in Japan (https://www.hpci-office.jp) under the Fugaku Junior Researchers Project (Project ID: hp210196). WR is supported by the Fundamental Research Funds for the Central Universities (2019XKQYMS75). JS is supported by the China Scholarship Council (202006420042).